\documentclass{misq} 
\usepackage{hyperref}
\hypersetup{
    colorlinks=true,
    linkcolor={blue},
    urlcolor={blue},
    citecolor=[rgb]{0,0.5,0.5},
}

\TheoremsNumberedThrough
\EquationsNumberedThrough

\usepackage{enumitem}
\usepackage{amsmath}
\usepackage{graphicx}
\usepackage{xcolor}
\usepackage[bottom]{footmisc}
\usepackage{booktabs}
\usepackage{boldline}
\usepackage{multirow}
\usepackage{fix-cm}
\usepackage{caption}
\usepackage{changepage}
\usepackage{algorithm}
\usepackage{algpseudocode}
\usepackage{makecell}

\newcommand{\vpara}[1]{\noindent\textbf{#1 }}

\begin{document}

\TITLE{
Guarding Organizations Against Malware Risk: A Novel Graph-Based Malware Detection Method 
}

\ARTICLEAUTHORS{%
\AUTHOR{Yinan Gao}
\AFF{School of Management, Fudan University, yngao25@m.fudan.edu.cn}
\AUTHOR{Jiarong Xu}
\AFF{School of Management, Fudan University, jiarongxu@fudan.edu.cn}
\AUTHOR{Xiaohang Zhao}
\AFF{School of Information Management and Engineering, Shanghai University of Finance and Economics, xiaohangzhao@mail.shufe.edu.cn}
\AUTHOR{Xiao Fang}
\AFF{Lerner College of Business and Economics, University of Delaware, xfang@udel.edu}
} 

\ABSTRACT{
\noindent The digitalization of business processes and IT infrastructures has expanded organizations’ exposure to cybersecurity risks, which have made cybersecurity an increasingly important research area in Information Systems (IS). Among these risks, malware has emerged as one of the most pervasive and destructive threats to organizational digital systems. To mitigate malware risks, byte-based machine learning (ML) methods are widely used to detect malware, but they remain vulnerable to malware exhibiting evasive behaviors. Such malware can evade detection by strategically manipulating raw bytes to confuse the detector while preserving its harmful functionality. Program graphs offer a promising alternative by representing software execution behavior rather than raw bytes, thereby reducing the influence of byte manipulations inserted into non-executed regions. Graph-based methods build on such program graphs, but existing approaches still face two limitations: they do not explicitly identify cohesive groups of basic blocks (i.e., short sequences of instructions) which jointly realize meaningful program behaviors, and they do not learn sufficiently expressive program graph representations for accurate malware detection. To address these limitations, we propose MalGuard, a graph-based malware detection method for supporting organizational malware risk management. MalGuard introduces two methodological innovations: an operational role identification approach and a program graph representation learning method. The former identifies these cohesive groups of basic blocks as operational roles, enabling the detector to capture program behaviors that may not be visible from isolated basic blocks. The latter learns expressive program graph representations by modeling interactions among operational roles, preserving sparse but critical malicious signals, and capturing the hierarchical structure of program graphs. Through extensive experiments on a real-world software dataset, we show that MalGuard outperforms state-of-the-art byte-based and graph-based malware detection methods and yields economic value by reducing the expected cost of undetected malware.
}


\maketitle

\newpage

\section{Introduction}
The digitalization of business processes and IT infrastructures has enabled organizations to improve operational efficiency and generate value from data-driven decision making~\citep{melville2004information, zheng2012business, mao2023personalized}. By moving more critical activities onto interconnected digital technologies, however, this transformation has expanded organizations’ exposure to cybersecurity risks~\citep{li2023information}.
{
Cybersecurity has therefore become a central theme in Information Systems (IS) research~\citep{cram2019seeing, abbasi2021phishing, samtani2022linking, clement2025dynamics}. In particular, IS scholars are increasingly adopting the computational design science paradigm~\citep{10.25300/MISQ/2017/41.1.E0, fang2025computational} to develop methods that help organizations mitigate cybersecurity risks, such as phishing threat \citep{abbasi2021phishing} and hacker exploits \citep{ampel2024creating}.
}
Among these risks, malware (malicious software) has emerged as one of the most pervasive and destructive threats to digital systems in organizations~\citep{kim2014differential, yu2009self}. Malware refers to software programs intentionally designed to infiltrate, disrupt, or damage computer systems~\citep{gandotra2014malware}.\footnote{In this paper, we focus on malware in the Windows Portable Executable (PE) file format, which accounts for over 97\% of all malware~\citep{ling2022malgraph}. We use the terms ``software'' and ``programs'' interchangeably.
}
The malware threat landscape remains large and active,
with tens of millions of new malware instances reported over the past 12 months.\footnote{See \href{https://portal.av-atlas.org/malware}{https://portal.av-atlas.org/malware} (last accessed on Jul 23, 2026).}
This growing threat has resulted in substantial financial losses for organizations. A prominent example is the SolarWinds supply-chain malware attack in 2020, which compromised nearly 18,000 organizations worldwide and caused estimated damages exceeding \$10 billion.\footnote{See \href{https://www.gao.gov/blog/solarwinds-cyberattack-demands-significant-federal-and-private-sector-response-infographic}{https://www.gao.gov/blog/solarwinds-cyberattack-demands-significant-federal-and-private-sector-response-infographic} (last accessed on Jul 23, 2026).}
Reflecting these consequences, research in IS has increasingly examined malware as both an organizational risk and a critical technical challenge for effective detection~\citep{guo2016impact,dey2026extortionality,ebrahimi2025radar}.

\begin{figure}[t]
    \centering
    \includegraphics[width=0.9\linewidth]{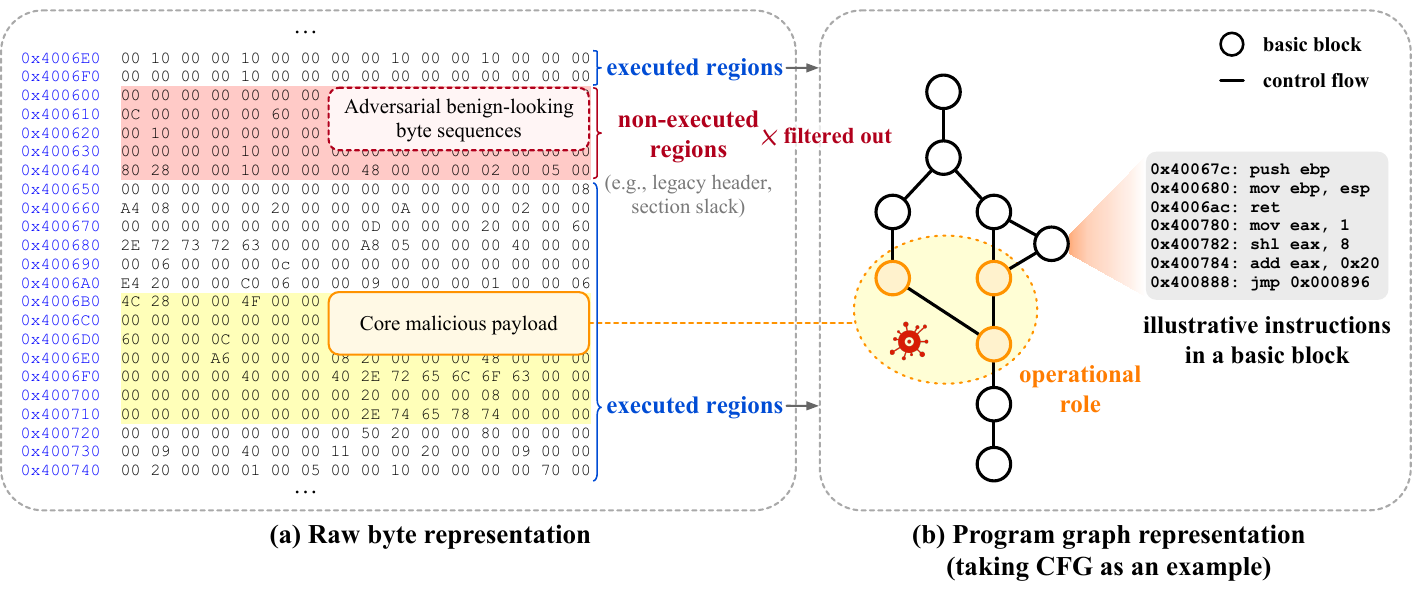}
\vspace{-0.1in}
\caption{Comparison of raw byte and program graph representations of malware with evasive behaviors.}
    \label{fig1}
\end{figure}

To mitigate malware risks, organizations increasingly rely on automated machine learning (ML) methods to detect malware before it can enter or operate within organizational environments. A large body of this research focuses on byte-based malware detection methods, which process the raw bytes of software as input and use ML models to learn byte-level patterns for distinguishing malware from benign software~\citep{raff2018malware,fleshman2018non,saha2024drsm}. However, such methods remain vulnerable to malware exhibiting evasive behaviors~\citep{afianian2019malware,d2020dissection}. In particular, adversaries can modify raw bytes so that malware evades detection without compromising its harmful functionality. This is often achieved by injecting adversarially crafted benign-looking byte sequences into non-executed regions while leaving the core executed malicious payload intact~\citep{kirat2014barecloud, d2020dissection, ebrahimi2025radar}. As illustrated in Figure~\ref{fig1}(a), these inserted adversarial benign-looking byte sequences (highlighted in red) can obscure the core malicious payload (highlighted in yellow) in the raw byte representation. Consequently, byte-based detection methods may be misled, resulting in the misclassification of evasive malware as benign~\citep{afianian2019malware, d2020dissection}. 
For organizations, such detection failures mean that malware may bypass cyber defenses and remain active within operational environments.

Graph-based malware detection methods address this challenge by shifting the representation of software from raw bytes to execution behavior modeled in program graphs. These program graphs are then fed into graph neural networks (GNNs) to learn software representations for detection
~\citep{yan2019classifying, ling2022malgraph, feng2024dawngnn, kargarnovin2024mal2gcn}. Figure~\ref{fig1} illustrates why this representation is more robust to malware with evasive behaviors.
Raw byte representations include both executed and non-executed byte regions, which means that byte-based detectors analyze the entire byte regions, including injected benign-looking byte sequences. In contrast, program graphs represent the program’s execution paths rather than its raw bytes. As shown in the resulting program graph of Figure~\ref{fig1}(b), ordinary code that is executed is mapped to regular graph nodes, and the core malicious payload is mapped to graph node(s) with malicious signals, whereas injected benign-looking bytes in non-executed regions have no corresponding graph nodes or edges and are filtered out. The program graph therefore preserves executed behavior while excluding non-executed injected bytes.
A common form of program graph is the control flow graph (CFG), where each node represents a basic block and each edge represents a possible control flow transition between basic blocks. 
A basic block is a short sequence of instructions that runs sequentially without internal branches or jumps; the inset in Figure~\ref{fig1}(b) provides an illustrative example of a basic block.
By abstracting execution flows rather than raw bytes, program graphs preserve executed behavior while excluding non-executed injected bytes, thereby providing robustness against evasive behaviors.

However, existing graph-based methods struggle to accurately detect malware when analyzing program graphs, leaving critical challenges unresolved. First, malicious behavior often does not reside in an individual basic block (node).
{As illustrated by the yellow blocks in Figure~\ref{fig1}(b), one block may reserve memory, another may prepare a request to the operating system, and a subsequent block may update a system setting. Examined in isolation, these blocks may appear routine: reserving memory, setting up a system request, or updating a system configuration. Yet when executed together along the control flow path, they can reveal a malicious privilege escalation operation, in which the malware attempts to obtain higher-level permissions than it is allowed to have.}
We refer to such a cohesive group of basic blocks that work together toward a shared computational objective as an \emph{operational role}~\citep{allen1970control, bruschi2006detecting}.
Identifying these operational roles is crucial for malware detection because it enables the detector to capture malicious behaviors that emerge only through coordinated execution. Existing graph-based methods typically learn representations from individual basic blocks or their local neighborhoods. Yet, operational roles are defined by shared computational objectives rather than by isolated basic blocks or fixed neighborhoods. Consequently, these methods fail to explicitly model operational roles, and  often overlook malicious signals that become apparent only when several routine-looking basic blocks are examined together.

Second, identifying operational roles alone is not sufficient; an effective detector must also learn expressive representations of program graphs for accurate malware detection. Addressing this challenge is nontrivial because program graphs exhibit three salient characteristics. 
The first characteristic is that operational roles may interact across the graph. 
{For example, one operational role may collect user credentials, while another may check these credentials to determine whether access to restricted system resources should be allowed.} Such interactions among roles are informative for understanding program behavior because they reveal higher-level workflows, such as authentication and access control. However, existing graph-based methods typically do not explicitly model interactions among operational roles.
{Furthermore, malicious signals for detecting malware are often sparse. A large program graph consists mostly of routine parts, such as initialization or simple computation, while only a small fraction carries malicious signals. Existing graph-based methods tend to combine information from different parts of the graph uniformly, both when GNNs aggregate information from neighboring nodes and when node representations are pooled to form the final software representation. In these steps, rare malicious signals may be mixed with routine neighboring signals or outweighed by representations of routine nodes. As a result, sparse but important malicious signals can become less visible to the detector.}
Besides, program graphs are structurally hierarchical, as software is organized at multiple levels. Basic blocks form the internal execution flow of each function, and functions further call one another to shape the behavior of the broader program.
A graph-based method therefore needs to model this hierarchy rather than treating the program  as a flat graph that ignores these levels.
Together, these characteristics call for an effective program graph representation learning method that can model complex interactions among operational roles, preserve sparse malicious signals as graph information is aggregated and pooled for detection, and capture the hierarchical structure of software.

In response to the above challenges, we propose {MalGuard}, a novel graph-based malware detection method to support malware risk management.
{MalGuard} is designed to overcome two critical limitations of existing graph-based ML methods: overlooking the operational roles in program graphs and falling short in learning expressive program graph representations for malware detection.
To achieve this, we first design an \emph{operational role identification approach} that uncovers cohesive groups of basic blocks functioning together toward shared computational objectives, i.e., operational roles.
By identifying these operational roles, our method captures node groups that can reflect meaningful program behaviors, including behaviors that may indicate malicious activity.
{Second, we propose a \emph{program graph representation learning method for malware detection}, featuring three key innovations: a coarsening layer to model complex interactions among operational roles, {attention-based message passing and gated pooling layers to preserve rare but critical malicious signals as graph information is aggregated and pooled for malware detection},
and a hierarchical learning framework to capture the hierarchical structure of program graphs and obtain the final software representation.}
Extensive experiments on a real-world software dataset collected for this study demonstrate that MalGuard outperforms state-of-the-art malware detection methods while yielding substantial economic value by reducing the financial cost of undetected malware.

\section{Literature Review}
In this section, we first review the broader IS cybersecurity literature and discuss malware threat. We then review existing malware detection methods and highlight the key novelties of our study.

\subsection{IS Cybersecurity Research and Malware Threat}
\label{sec:2.1}

{Our study is situated in IS cybersecurity research, an important area that examines cybersecurity as a critical organizational and technical challenge.
IS cybersecurity research can be broadly organized into three categories.
One stream examines the behavioral dimension of cybersecurity, focusing on how individuals and organizations perceive and respond to cybersecurity threats~\citep{bulgurcu2010information, moody2018toward, cram2019seeing}.
A second stream uses economic modeling to examine how cybersecurity risks influence firms’ market outcomes and economic consequences, thereby generating managerial insights for firms’ cybersecurity risk management~\citep{clement2025dynamics, temizkan2017software, august2022economics}.
A third stream, more relevant to our study, adopts the computational design science paradigm to develop technical methods that help organizations mitigate cybersecurity risks, {such as phishing threat, hacker exploits, and cyber threats associated with hacker assets}~\citep{abbasi2021phishing, samtani2022linking, ampel2024creating}. For example, \citet{abbasi2021phishing} address phishing as a cybersecurity threat by proposing a phishing funnel model to predict user susceptibility to phishing websites.
{
\citet{samtani2022linking} and \citet{ampel2024creating} both focus on hacker exploits as cybersecurity threats and develop technical methods to analyze and label exploits for proactive cyber defense.
} In addition, \citet{ebrahimi2022cross} study cyber threats associated with hacker assets, {such as hacking tools and malicious source code shared on the dark web}, and introduce cross-lingual cybersecurity analytics for detecting such assets across international dark web platforms.
These studies demonstrate the value of computational design science IS research in responding to different cyber threats.
}

Among the cybersecurity risks, malware (malicious software) creates a critical challenge for organizational risk management. Malware is intentionally developed by cybercriminals to achieve harmful objectives, typically targeting enterprise computer systems and leading to substantial economic and societal damage~\citep{kolter2004learning}.
Given these profound implications, IS scholars have increasingly focused on the economic impacts and detection mechanisms of malware.
Within this domain, one line of IS research focuses on the economic and organizational implications of malware, aiming to explain how malware affects market competition and organizational processes~\citep{kim2014differential, guo2016impact, august2022economics}. For instance, \citet{kim2014differential} study the malware resolution process of an antivirus software provider and show how prior experience affects the efficiency of resolving malware problems. \citet{guo2016impact} further analyze how network structures dictate the propagation speed and scale of malware. In addition, ransomware has received increasing attention as a major form of malware risk, with recent studies examining its implications for risk interdependence and externalities across firms~\citep{august2022economics,dey2026extortionality}.
Another line of IS research focuses on developing technical methods for malware detection~\citep{ebrahimi2025radar}. \citet{ebrahimi2025radar} propose the RADAR framework, which utilizes deep reinforcement learning to develop an adversarially robust malware detection method.
Together, these studies establish malware as an important cybersecurity topic in IS research and highlight the need for effective technical artifacts to support organizational malware risk management.

To ground this research in a prevalent organizational malware threat, we focus on Portable Executable (PE) malware, which has emerged as one of the most common and persistent cybersecurity threats~\citep{ye2017survey}.
PE malware targets programs in the PE format, the standard file format for Windows applications~\citep{microsoftpeformat}. Because Windows remains widely used in consumer and enterprise computing, PE software represents a high-value target for attackers. 
In addition, the structured and extensible nature of the PE format provides attackers with multiple opportunities to manipulate file contents~\citep{pietrek2002depth}. 
These characteristics make PE malware a persistent threat to organizations, supporting a wide range of malicious activities such as data theft, system disruption, and ransomware deployment~\citep{august2022economics,mcintosh2024ransomware,dey2026extortionality}. 
More importantly, the challenge of PE malware is further amplified by its increasing use of \emph{evasive behaviors}. Malware with evasive behaviors deliberately modifies its byte file to evade detection while preserving harmful functionality~\citep{christodorescu2003static,kirat2015malgene,d2020dissection,ling2023adversarial}. 
These evasion techniques are designed to mislead ML-based detectors by injecting adversarial or benign-looking byte sequences into non-executed regions of the file, such as DOS headers, slack space, or appended regions~\citep{demetrio2021adversarial,suciu2019exploring,kreuk2018deceiving,nisi2021lost}. As illustrated in Figure~\ref{fig1}(a), such injected bytes (in red) can obscure the core malicious payload (in yellow) in the raw byte representation while leaving the executed malicious behavior intact. These evasive behaviors make malware increasingly difficult to detect, posing a serious threat to malware detection systems.

Our study contributes to IS cybersecurity research by proposing a graph-based malware detection method that provides inherent robustness to malware with evasive behaviors and more accurately identifies malware by effectively learning from program structures and malicious signals, thereby supporting organizational malware risk management.

\subsection{Malware Detection Methods}
\label{sec:2.2}
To identify and mitigate malware threat, a variety of malware detection methods have been developed. These methods can be broadly categorized into four main types based on how they represent software for detection:
signature-based methods, feature-based ML methods, byte-based ML methods, and graph-based ML methods. Among these, signature-based  and feature-based methods are traditional detection approaches.
Signature-based methods match predefined patterns from known malware~\citep{sung2004static}, while feature-based ML methods train classifiers on handcrafted software features~\citep{mohaisen2015amal,anderson2018ember}; however, both approaches struggle to detect malware with evasive behaviors, and their reliance on fixed signatures or handcrafted features limits their ability to capture more complex behavioral patterns of malware~\citep{ye2007imds,anderson2018ember}.
Recent malware detection research has increasingly shifted toward byte-based and graph-based ML methods. Accordingly, we focus on these two approaches and review them in detail below.

\vpara{Byte-based ML malware detection.}
Byte-based methods operate directly on the raw byte sequences of programs and learn malware-related patterns from the original bytes. These methods commonly adopt convolutional neural networks (CNNs) to capture local byte-level patterns in executables~\citep{yan2018detecting}.
A prominent example is MalConv~\citep{raff2018malware}, which embeds each byte sequence into a continuous vector space, applies gated convolutional layers to learn local patterns, and uses temporal max-pooling to extract the most salient features across the file. MalConv demonstrates strong effectiveness in malware detection and remains a state-of-the-art benchmark for byte-based approaches.

Despite their effectiveness in detecting conventional malware, MalConv and other byte-based models remain vulnerable to malware exhibiting evasive behaviors~\citep{ling2023adversarial}.
This vulnerability arises because these methods process the entire PE file, including non-executed regions where adversarial perturbations or benign-looking byte sequences can be injected~\citep{kreuk2018deceiving, demetrio2019explaining, demetrio2021adversarial, suciu2019exploring, nisi2021lost}.
To mitigate this issue, several extensions of MalConv have been proposed.
For instance, \citet{fleshman2018non} impose non-negative constraints on the final layer’s weights to reduce the model’s sensitivity to the manipulated regions, preventing injected bytes from undermining malware predictions.
\citet{saha2024drsm} propose an ensemble-based strategy DRSM, which divides the input byte sequence into multiple segments, applies a pre-trained MalConv model to each segment independently, and aggregates the resulting predictions through majority voting. This design improves robustness by reducing over-reliance on any single region of the file.
Although these methods alleviate the impact of evasive behaviors, the use of constraints, region-wise ensemble mechanisms, or adversarial training may compromise detection accuracy on conventional malware~\citep{saha2024drsm}.

\vpara{Graph-based ML malware detection.}
Graph neural networks (GNNs) have become a widely used framework for learning representations from graph-structured data~\citep{wu2020comprehensive,feng2017efficient}. 
Using this capability, graph-based ML methods detect malware by representing software as program graphs, which are then fed into GNNs for detection. Compared to byte-based methods, these methods offer two key advantages.
First, by encoding control structures and function call interactions in program graphs, graph-based ML approaches can better capture a software instance’s underlying execution logic~\citep{reps1998program, alfred2007compilers}.
Second, these approaches are inherently more robust to malware exhibiting evasive behaviors. Because program graphs model the execution logic of software, benign-looking byte sequences injected into non-executed regions are excluded from the graph and thus have limited influence on detection outcomes~\citep{bilot2024survey}.
Figure~\ref{fig1}(b) shows a program graph of malware with evasive behaviors, which preserves the executed malicious functionality (in yellow) while naturally removing the adversarial benign-looking byte sequences that are never executed.

Several recent studies have advanced graph-based ML methods to model program graphs for malware detection. For example, \citet{yan2019classifying} propose MAGIC, which represents software as a control flow graph (CFG) to explicitly capture execution paths.
A GNN is then applied to learn structural patterns from the CFG.
\citet{chen2023deepcall} introduce DeepCall, which constructs a call graph (CG) from each software, where nodes represent functions and edges denote calling relationships, and applies a GNN to learn from the CG.
Furthermore, \citet{kargarnovin2024mal2gcn} explore CGs for malware detection and impose a non-negative weight constraint to enhance the robustness of GNNs against evasive behaviors, inspired by earlier non-negative techniques for byte-based methods~\citep{fleshman2018non}.
Nevertheless, these approaches only focus on a single structural view of program graph, either CFG or CG.
To overcome this limitation, \citet{ling2022malgraph} propose MalGraph, which integrates CFGs and CG into a unified hierarchical structure and employs a hierarchical framework to learn from the structure.

However, existing graph-based ML methods do not fully exploit program graphs for malware detection. 
\emph{First}, meaningful behaviors in CFGs often emerge not from individual nodes but from groups of nodes working together toward a shared computational objective, which we refer to as an operational role~\citep{balsamo2004model}. 
These roles may directly correspond to malicious intent, such as privilege escalation. 
Yet existing graph-based methods typically learn from individual nodes or from their local neighborhoods, without explicitly identifying operational roles. 
This creates a mismatch because an individual basic block often captures only a short sequence of instructions rather than a meaningful behavior, while local neighborhoods do not group nodes according to whether they jointly carry out the same computational objective. 
As a result, these methods may miss malicious behavior that emerges only when several routine-looking blocks are examined together.
\emph{Second}, existing methods still fall short in learning expressive program graph representations for malware detection. 
Effective detection requires software representations that can jointly capture complex interactions among operational roles, preserve sparse but critical malicious evidence during GNN aggregation and pooling for detection, and model the hierarchical structure between CFGs and the CG. 
Existing GNN architectures are limited in this context, which underscores the need for a more effective graph representation learning approach for malware detection.

\begin{table}[t]
    \centering
    \caption{Comparison of our proposed method with existing malware detection methods.}
    \label{tab:gap}
    \resizebox{\textwidth}{!}{
    \renewcommand{\arraystretch}{1}
    \begin{tabular}{ll|
                >{\centering\arraybackslash}m{3cm} 
                >{\centering\arraybackslash}m{3cm} 
                >{\centering\arraybackslash}m{3cm} 
                >{\centering\arraybackslash}m{3cm}
                >{\centering\arraybackslash}m{3cm}}
        \toprule
        
        {\textbf{Category}} &
        {\textbf{Methods}} & 
        \small{\textbf{Robust to evasive behaviors}} & 
        \small{\textbf{Operational role identification}} & 
        \small{\textbf{Modeling role interactions}} &
        \small{\textbf{Preserving sparse malicious signals}} &
        \small{\textbf{Modeling hierarchical structure}}
        \\
        \midrule
        
        \multirow{3}{*}{Byte-based} 
        & MalConv~\citep{raff2018malware}
        & $\times$ & $\times$ & $\times$ & $\times$ & $\times$ \\
        & NonNeg~\citep{fleshman2018non}
        & $\checkmark$ & $\times$ & $\times$ & $\checkmark$ & $\times$ \\
        & DRSM~\citep{saha2024drsm}
        & $\checkmark$ & $\times$ & $\times$ & $\checkmark$ & $\times$ \\
        \midrule

        \multirow{4}{*}{Graph-based} 
        & MAGIC~\citep{yan2019classifying}
        & $\checkmark$ & $\times$ & $\times$ & $\times$ & $\times$ \\
        & MalGraph~\citep{ling2022malgraph}
        & $\checkmark$ & $\times$ & $\times$ & $\times$ & $\checkmark$ \\  
        & DeepCall~\citep{chen2023deepcall}
        & $\checkmark$ & $\times$ & $\times$ & $\times$ & $\times$ \\
        & Mal2GCN~\citep{kargarnovin2024mal2gcn}
        & $\checkmark$ & $\times$ & $\times$ & $\checkmark$ & $\times$ \\
        \midrule
        
        \multicolumn{2}{c|}{\textbf{Ours}}
        & $\checkmark$ & $\checkmark$ & $\checkmark$ & $\checkmark$ & $\checkmark$ \\
        \bottomrule
    \end{tabular}}
\end{table}

\vpara{Research gaps.}
Our literature review suggests that mainstream malware detection methods, including byte-based and graph-based approaches, suffer from inherent limitations, as summarized in Table~\ref{tab:gap}.
Byte-based ML methods treat programs as raw bytes, overlooking the execution logic that is critical for understanding program behavior. Moreover, these methods are vulnerable when detecting malware exhibiting evasive behaviors because they rely on raw byte patterns that adversaries can strategically manipulate while preserving malware’s harmful functionality. 
Graph-based ML methods are more robust to evasive behaviors because they shift the basis of detection from such surface-level artifacts to execution behavior modeled in program graphs. Since program graphs are constructed from possible execution paths, adversarial manipulations inserted into non-executed regions have limited impact on the resulting representation. However, important research gaps remain.
\emph{First}, existing graph-based methods do not explicitly identify operational roles, which are groups of basic blocks that jointly carry out a shared computational objective. This limitation is important because malicious behavior may not be visible in individual blocks or their local neighborhoods, but may become apparent only when several routine-looking blocks are examined together as part of the same operational role.
\emph{Second}, existing graph-based methods leave the challenge of learning expressive program representations for malware detection. Effective representations must simultaneously capture interactions among operational roles, preserve sparse but critical malicious signals during aggregation and pooling, and model hierarchical program structures. Existing architectures are not designed to satisfy these requirements simultaneously.

To address these gaps, we propose two methodological components. First, we develop an {operational role identification approach} that infers semantically cohesive groups of basic blocks corresponding to shared computational objectives. Second, we propose a {program graph representation learning method for malware detection} that can simultaneously model interactions among operational roles, preserve sparse malicious evidence for accurate malware detection, and capture hierarchical program structure. Together, these two components constitute the key methodological contributions of this study.

\section{Problem Formulation and Preliminaries}
\label{sec:3}
In this section, we first formally define the malware detection problem. We then introduce preliminaries on program graphs, which serve as the input of our proposed malware detection method.

\subsection{Problem Definition}
\label{sec:3.1}
Consider a set of software instances denoted by $\mathcal{R}$, where each instance $r_i \in \mathcal{R}$ is assigned a binary label $y_i \in \{0, 1\}$.
Specifically, $y_i = 0$ indicates that the software is benign, whereas $y_i = 1$ denotes that it is malicious.
Formally, we define the research problem as follows.

\begin{definition}[Malware Detection Problem]
Let $\mathcal{R} = \{ r_i \}$ denote a set of software instances, where each instance belongs to one of two classes: benign or malicious.
The objective of malware detection is to learn a classification function
\begin{equation}
    f: r_i \rightarrow y_i, \quad y_i \in \{0,1\},
    \label{eq:obj}
\end{equation}
that can accurately identify malicious instances ($y_i = 1$), including those with evasive behaviors.
\end{definition}

\subsection{Hierarchical Program Graph}
\label{sec:3.2}
Representing software in a form that is inherently robust to evasive behaviors is a prerequisite for addressing the malware detection problem. Program graphs provide such a representation by abstracting a program’s potential execution logic. Because program graphs are constructed from possible execution paths rather than raw bytes, manipulations inserted into non-executed regions, a common evasion strategy, have limited impact on the resulting graph~\citep{peng2025evading}.

\begin{figure}[t]
\centering
\makebox[\textwidth][c]{
    \includegraphics[width=0.8\textwidth]{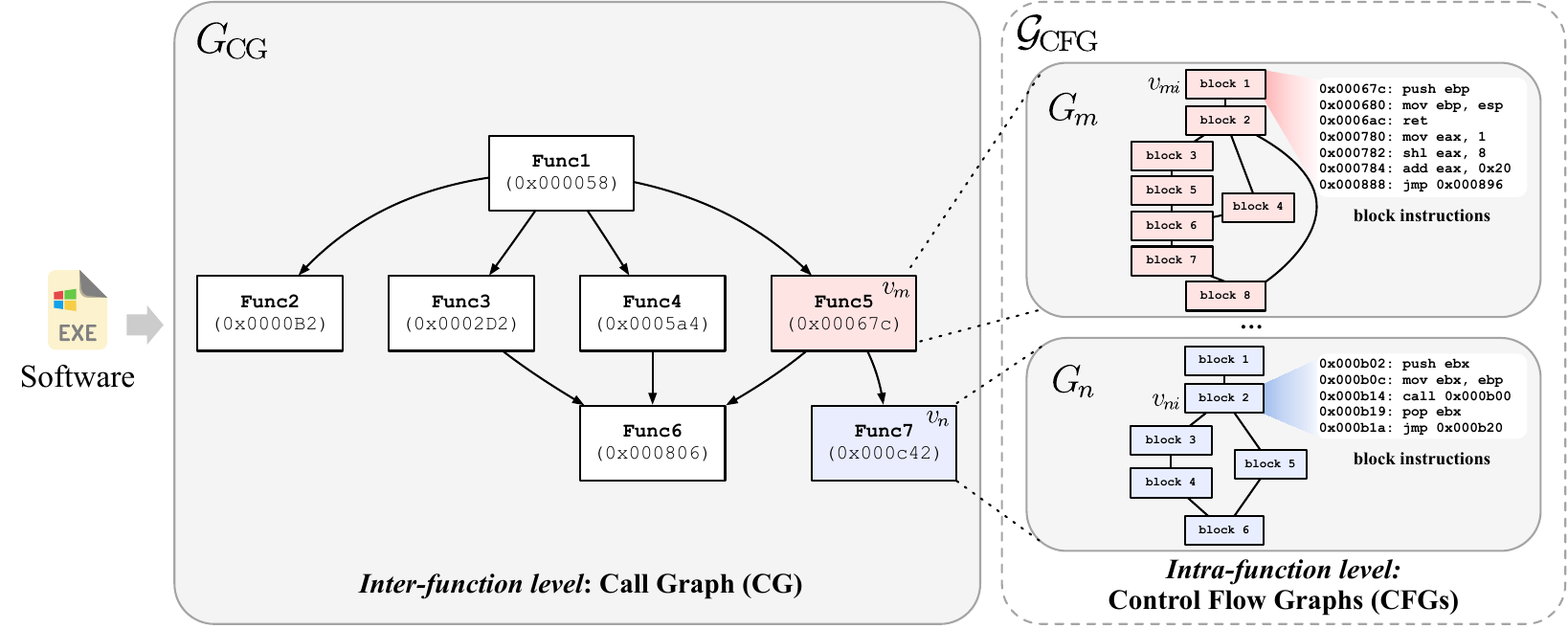}
}
\caption{Illustration of the hierarchical program graph for a software instance.}
\label{fig2}
\end{figure}

Program execution is complex because software is organized hierarchically. 
Software is not simply a flat sequence of instructions; instead, a software program consists of multiple functions, and functions may call one another during execution. These inter-function relationships shape the behavior of the overall program and are represented by a Call Graph (CG). At the same time, within each function, instructions are grouped into multiple basic blocks, and the intra-function control flow structure among basic blocks is represented by a CFG.
Figure~\ref{fig2} illustrates such a hierarchical structure in a program graph for an example software instance.

We therefore represent each software instance as a \emph{hierarchical program graph} that combines CFGs for individual functions with a CG connecting these functions. In this hierarchy, CFGs constitute the intra-function (lower) level and the CG constitutes the inter-function (higher) level. This representation captures both how execution proceeds within each function and how functions interact across the program. 

Formally, let $G_{\text{CG}}$ denote the CG among the functions in a software instance, with $V_{\text{CG}}$ denoting its set of function nodes and $|V_{\text{CG}}|$ the number of functions. Let $\mathcal{G}_{\text{CFG}}=\{G_1,G_2,\ldots,G_{|V_{\text{CG}}|}\}$ denote the collection of CFGs, where $G_m$ is the CFG corresponding to function node $v_m \in V_{\text{CG}}$.
The hierarchical program graph is defined as
\begin{equation}
    \mathcal{H} = (G_{\text{CG}}, \mathcal{G}_{\text{CFG}}).
\end{equation}
Next, we detail the construction procedure of a hierarchical program graph.

At the inter-function level, the software is transformed into a CG, which captures interactions between functions, with nodes representing functions and edges indicating call instructions between caller functions and their corresponding callees~\citep{ryder1979constructing}.
We define the CG as follows:
\begin{definition}
[Call Graph (CG)]
Given a software instance, its CG is defined as $G_{\text{CG}} = (V_{\text{CG}}, E_{\text{CG}})$. Here, $V_{\text{CG}}$ is the set of nodes, with each node $v_{m} \in V_{\text{CG}}$ representing a function contained in the software. 
$E_{\text{CG}}$ is the set of edges, with each edge $(v_m, v_n) \in E_{\text{CG}}$ denoting a function call between the caller function $v_m$ and the callee function $v_n$.
\end{definition}

At the intra-function level, each function is further represented as a CFG.
A CFG specifies the internal control flow of a function, where nodes correspond to basic blocks and edges represent control flow transitions between them~\citep{ryder1979constructing}.
A basic block is a straight-line sequence of instructions that executes sequentially from a single entry point to a single exit point, without internal control flow transfers such as branches, jumps, or returns.
Each basic block is further enriched with statistical attributes extracted from its instructions.
CFG can be defined as follows:

\begin{definition}
[Control Flow Graph (CFG)]
Given the function represented by node $v_m \in V_{\text{CG}}$, its CFG is defined as 
$G_m = (V_m, A_m, X_m)$, where $V_m$ is the set of basic block nodes, 
$A_m \in \{0,1\}^{|V_m|\times |V_m|}$ is the adjacency matrix, and 
$X_m \in \mathbb{R}^{|V_m| \times d_x}$ is the node attribute matrix. 
Each node $v_{mi} \in V_m$ represents the $i$-th basic block within function $m$, and $|V_m|$ is the total number of basic blocks within the function. 
For the adjacency matrix, $(A_m)_{ij}=1$ indicates the existence of a control flow from basic block $v_{mi}$ to basic block $v_{mj}$, and $(A_m)_{ij}=0$ otherwise.
The $i$-th row of the node attribute matrix, $(X_m)_i$, encodes statistical attributes derived from the instruction content of basic block $v_{mi}$, including the total number of instructions, the counts of different instruction types, constant values, and other statistical characteristics.\footnote{When focusing on a single CFG and no ambiguity arises, we simplify the notation to $G=(V,A,X)$ and write $v_i\in V$, $A_{ij}$, and $X_i$ for basic block nodes, adjacency entries, and node attributes, respectively.}
\end{definition}

The constructed hierarchical program graph provides a unified abstraction of software execution behavior by capturing both inter-function interactions and intra-function control flows. 
By design, evasive manipulations inserted into regions that do not appear on execution paths have limited impact on the resulting hierarchical program graph, making it a suitable input for the method proposed in the next section.

\section{Proposed Method}
\begin{figure}[t]
\centering
\makebox[\textwidth][c]{
    \includegraphics[width=1\textwidth]{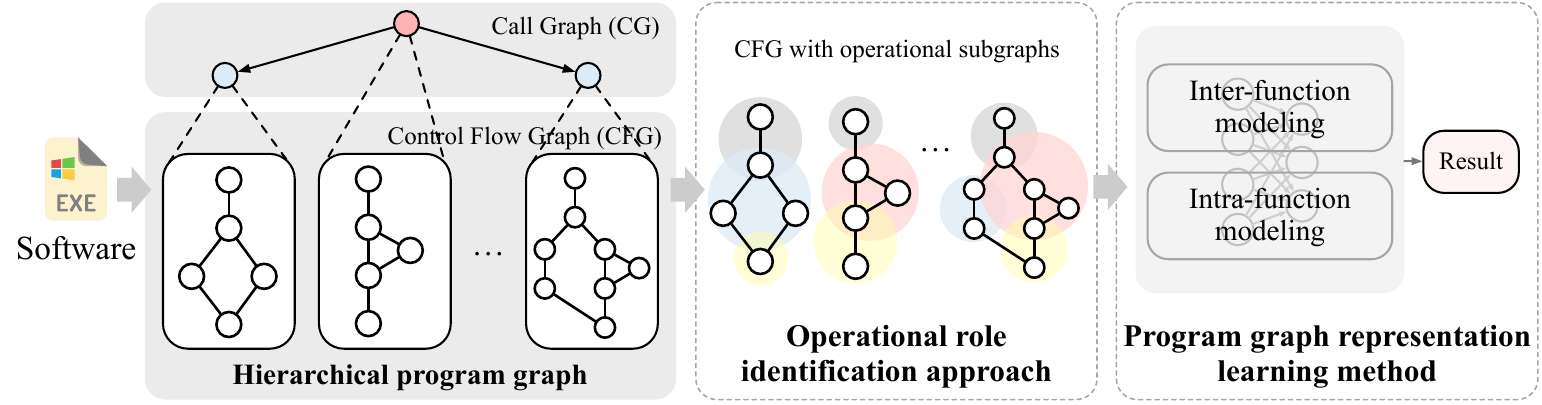}
}
\caption{Framework overview of MalGuard.}
\label{fig3}
\end{figure}

In this section, we introduce MalGuard, a novel graph-based method for addressing the malware detection problem.
MalGuard builds on hierarchical program graphs, which represent software execution behavior and provide robustness against malware with evasive behaviors.
MalGuard introduces two methodological innovations. First, an {operational role identification} approach infers latent operational roles in each CFG and extracts the corresponding operational subgraphs (Section~\ref{sec:4.1}). Second, a {program graph representation learning method for malware detection} learns software representations for accurate malware detection by modeling interactions among operational roles, preserving sparse malicious signals, and capturing the hierarchical structure of program graphs (Section~\ref{sec:4.2}). Section~\ref{sec:4.3} summarizes the training algorithm.
Figure~\ref{fig3} provides a framework overview of the proposed method.

\subsection{Operational Role Identification}
\label{sec:4.1}
A salient characteristic of the hierarchical program graph is that within each CFG, meaningful execution behaviors often arise not from individual basic blocks (nodes) but from groups of basic blocks working together toward a shared computational objective~\citep{allen1970control, bruschi2006detecting}.
We refer to such a semantically cohesive group of basic blocks as having a shared \emph{operational role} (short for \emph{role}).

Unlike a single basic block that only performs elementary operations and thus contributes just a fragment of an overall behavior, this collaborative structure among basic blocks with a shared role captures a more holistic execution behavior that emerges from the joint activity of multiple blocks. More importantly, it is particularly crucial in malware detection.
Malicious behaviors are typically realized through the coordinated execution of interrelated blocks~\citep{cesare2010classification}. 
Therefore, identifying operational roles of basic blocks is essential for accurately modeling CFGs and uncovering malicious patterns. 
However, these roles are not directly observable or labeled in practice, and manual identification is both labor-intensive and impractical at scale. 
To address this, we propose a novel {operational role identification} module that treats operational roles as latent variables, enabling them to be inferred from the observed CFG without relying on explicit supervision. We first overview the module architecture and then elaborate each component in turn.

\subsubsection{Overall Architecture}
{Inspired by latent variable approaches that infer hidden constructs from observable data~\citep{zheng2010causal, wang2024m3rec}, we represent operational roles as latent variables associated with program nodes.} Formally, given a CFG $G = (V, A, X)$, each basic block (node) $v_i \in V$ is associated with a latent variable $\mathbf{z}_i \in \mathbb{R}^K$ encoding its probabilistic affiliation with one or more operational roles.
Each dimension $\mathbf{z}_{ik}$ corresponds to the degree to which $v_i$ is associated with the $k$-th operational role. 
A notable feature of CFG is that a single basic block may contribute to multiple roles simultaneously. For example, in a ransomware, a single basic block might simultaneously encrypt a user’s file (a malicious role) and check whether it is the last file in a folder (a control role)~\citep{alfred2007compilers}.
To accommodate this multi-role assignment of a basic block, we model $\mathbf{z}_i$ using a \emph{multivariate logit-normal distribution}~\citep{mead1965generalised, atchison1980logistic, aitchison1982statistical}, ensuring that each element $\mathbf{z}_{ik}$ lies within the range $(0, 1)$.
This probabilistic formulation supports soft yet interpretable role assignments and allows each block to participate in multiple roles to varying extents.\footnote{Compared to conventional Gaussian priors commonly used in existing works~\citep{kipf2016variational, grover2019graphite}, the multivariate logit-normal distribution is better suited for reflecting latent roles in real-world software behaviors.}

\begin{figure}[t]
\centering
\makebox[\textwidth][c]{
    \includegraphics[width=0.7\textwidth]{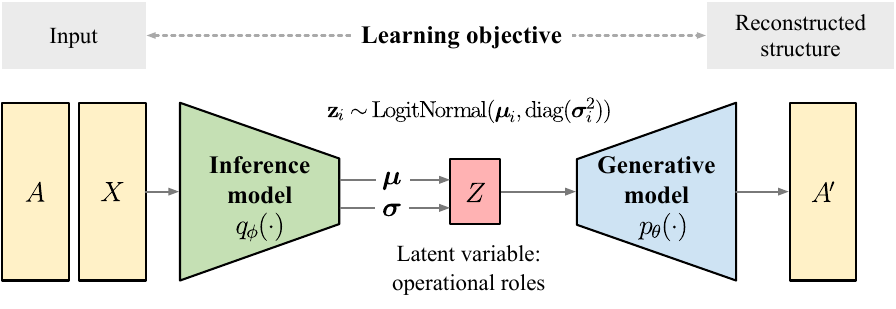}
}
\caption{The architecture of the operational role identification module.}
\label{fig4}
\end{figure}

To infer the latent operational roles from observed CFG, we adopt a variational inference-generative architecture as illustrated by Figure~\ref{fig4}: the inference model maps each basic block to a distribution over a low-dimensional latent space, capturing its probabilistic association with latent roles; the generative model then attempts to reconstruct the original CFG structure based on the sampled latent role assignments, encouraging the learned roles to preserve meaningful execution-related semantics. This reconstruction-based training objective allows the model to learn latent roles that are operationally meaningful and aligned with execution semantics. 
In the following, we detail the design of the inference model, the generative model, and the training objective.

\subsubsection{Inference Model}
The inference model aims to infer a latent variable of each basic block that captures its association with multiple latent operational roles. Specifically, the inference model maps the adjacency matrix $A$ and node attribute matrix $X$ of the CFG into a variational posterior distribution $q_\phi(Z|A,X)$ where $Z = \{ \mathbf{z}_i\}_{i=1}^{|V|}$ is the set of latent variables.
Following prior works~\citep{kipf2016variational, grover2019graphite, li2020dirichlet}, we adopt a mean-field approximation for the variational distribution by assuming that the latent variables are conditionally independent given the input graph.
In accordance with the aforementioned logit-normal design of each latent vector $\mathbf{z}_i$, we model each $q_\phi(\mathbf{z}_i \mid A, X)$ as a multivariate logit-normal distribution:
\begin{gather}
    q_\phi(Z|A,X) = \prod_{i=1}^{|V|} q_{\phi}(\mathbf{z}_i|A,X),\\
    q_{\phi}(\mathbf{z}_i|A,X) = \text{LogitNormal}(\boldsymbol{\mu}_i,\Sigma_i), \ \Sigma_i=\operatorname{diag}(\boldsymbol{\sigma}_i^2),
\end{gather}
where $\text{LogitNormal}(\boldsymbol{\mu}_i,\Sigma_i)$ is the density function of the logit-normal distribution specified by the mean vector $\boldsymbol{\mu}_i  \in \mathbb{R}^K$ and the covariance matrix $\Sigma_i \in\mathbb{R}^{K \times K}$, with the latter being a diagonal matrix filled by the element-wise square of the standard deviation vector $\boldsymbol{\sigma}_i \in \mathbb{R}^K$.
This formulation guarantees that $\mathbf{z}_{ik} \in (0,1)$ for $k=1,2,...K$. 
In plain words, $\boldsymbol{\mu}_i$ quantifies the expected logit vector of role strengths for node $v_i$, while $\boldsymbol{\sigma}_i$ captures the uncertainty in these estimates across different roles.
To obtain $\boldsymbol{\mu}_i$ and $\boldsymbol{\sigma}_i$, we apply a GNN that encodes both structural information $A$ and node attributes $X$ simultaneously:
\begin{equation}
    \boldsymbol{\mu}, \boldsymbol{\sigma} = \text{GNN}_\phi(A,X),
    \label{eq:mu}
\end{equation}
where $\text{GNN}_\phi$ is a GNN layer specified by learnable parameters $\phi$. This layer outputs a matrix of size $|V| \times 2K$, with the first $K$ columns extracted as matrix $\boldsymbol{\mu} \in \mathbb{R}^{|V| \times K}$, while the remaining $K$ columns as $\boldsymbol{\sigma} \in \mathbb{R}^{|V| \times K}$. We then treat the $i$-th row of $\boldsymbol{\mu}$ and $\boldsymbol{\sigma}$ as $\boldsymbol{\mu}_i$ and $\boldsymbol{\sigma}_i$ respectively.

To enable gradient-based optimization, we adopt the reparameterization trick~\citep{kingma2014auto, rezende2014stochastic} to draw instances from $q_{\phi}(\mathbf{z}_i|A,X)$ in a way that is differentiable with respect to learnable parameters.
Specifically, once obtaining $\boldsymbol{\mu}_i$ and $\boldsymbol{\sigma}_i$, we sample the latent variable $\mathbf{z}_i$ as follows:
\begin{gather}
    \widetilde{\mathbf{z}}_i = \boldsymbol{\mu}_i + \boldsymbol{\sigma}_i \odot \epsilon_i, \ \ \text{where}\ \epsilon_i \sim \mathcal{N}(0, I),  \\
\mathbf{z}_{ik} = \text{sigmoid}(\widetilde{\mathbf{z}}_{ik}), \ \ k=1,2,...,K,
\label{eq:z}
\end{gather}
where $\odot$ denotes element-wise multiplication, $\epsilon_i \in \mathbb{R}^{K}$ is a random noise vector drawn from the standard multivariate normal distribution $\mathcal{N}(\boldsymbol{0}, I)$, and $\text{sigmoid}(z)=1/(1+\exp(-z))$.
The sigmoid function ensures $\mathbf{z}_{ik} \in (0, 1)$, allowing it to be interpreted as the probability of role $k$ being assigned to node $v_i$.
This reparameterization-based sampling procedure faithfully instantiates a distribution that is multivariate logit-normal.
The resulting latent variables $Z$ then serve as a semantically meaningful and probabilistically grounded representation of the operational roles played by the basic blocks of the given CFG.

\subsubsection{Generative Model}
To guide the learning of latent operational roles toward reflecting meaningful execution behavior, we design a generative model $p_\theta(A|Z)$ that reconstructs the original CFG structure based on the latent roles of its constituent basic blocks. The underlying intuition is that accurate reconstruction requires the latent roles to encode execution-related semantics.
In particular, this generative model is responsible for generating the adjacency matrix $A$, which encodes the presence or absence of edges (i.e., control flow relationships), based solely on the latent role assignment information $Z$. 
Since the CFG structure is composed entirely of edges between basic blocks, generating the graph reduces to deciding, for every node pair $(v_i, v_j)$, whether $A_{ij}$ equals one or zero, given their corresponding nodes’ latent roles $\mathbf{z}_i$ and $\mathbf{z}_j$. Therefore, the generative model defines the conditional probability of the full adjacency matrix as a product of the probabilities of individual matrix entries:
\begin{equation}
p_\theta(A|Z) = \prod_{A_{ij}=1} p_\theta(A_{ij} = 1 \mid \mathbf{z}_i, \mathbf{z}_j) \prod_{A_{ij}=0} p_\theta(A_{ij} = 0 \mid \mathbf{z}_i, \mathbf{z}_j).
\end{equation}

To instantiate this model, we need to specify how the probability of edge presence or absence (i.e., $p_\theta(A_{ij} = 1 \mid \mathbf{z}_i, \mathbf{z}_j)$ and $p_\theta(A_{ij} = 0 \mid \mathbf{z}_i, \mathbf{z}_j)$) is determined from the latent roles. Conventional edge generation strategies assume that similar latent variables should imply high likelihood of edge formation~\citep{kipf2016variational, grover2019graphite, li2020dirichlet}. However, such an assumption does not necessarily hold in the case of CFGs. A distinctive property of CFGs is that nodes with similar or even identical latent roles may be located far apart in the graph without direct connections~\citep{musgrave2024empirical}. This is common for reusable routines such as error-handling or control-check blocks, which are invoked repeatedly at different points during program execution~\citep{haq2021survey}. For instance, an error-handling routine may occur in multiple branches of a software yet remain disconnected in the CFG. This property challenges the validity of conventional edge generation strategies in CFGs. 

As such, we require a more flexible edge generation strategy, one that incorporates role similarity as a useful signal while still allowing for the absence of edges between nodes with similar or identical roles. 
In this regard, we introduce a novel \emph{edge generation strategy with a learnable gate}, which is explicitly designed to capture the structural particularity of CFGs.
Specifically, for edge presence, we follow the intuition that nodes with highly similar roles are more likely to be connected.
Edge absence, however, is more nuanced: two nodes may be disconnected either because they serve different roles, or because, despite having similar roles, they lie on distinct branches in the control flow and therefore are not directly connected in the CFG. 
To model these two cases of edge absence, we introduce a learnable gate $M$ that adaptively determines whether an absent edge should be explained by role dissimilarity or by the fact that two nodes with similar roles are in distinct branches in the control flow.
Formally, the edge generation likelihoods are defined as:
\begin{gather}
\begin{cases}
p_\theta(A_{ij} = 1 \mid \mathbf{z}_i, \mathbf{z}_j) = \text{sigmoid} \bigl(\text{sim}(\mathbf{z}_i, \mathbf{z}_j)\bigr), \\
p_\theta(A_{ij} = 0 \mid \mathbf{z}_i, \mathbf{z}_j) = \text{sigmoid} \bigl(M_{ij}(1 - \text{sim}(\mathbf{z}_i, \mathbf{z}_j)) + (1 - M_{ij}) C \bigr), 
\end{cases} \\
M_{ij} = \text{sigmoid}\bigl(\text{MLP}_\theta(\mathbf{z}_i, \mathbf{z}_j)\bigr),
\end{gather}
where $\text{sim}(\mathbf{z}_i, \mathbf{z}_j)= \mathbf{z}_i^\top \mathbf{z}_j / (\|\mathbf{z}_i\|_2 \|\mathbf{z}_j\|_2)$ denotes the similarity between the latent roles $\mathbf{z}_i$ and $\mathbf{z}_j$, and $\text{MLP}_\theta$ is an MLP (Multi-Layer Perceptron) layer specified by learnable parameters $\theta$. 
The constant $C$ provides a fallback likelihood for edge absence 
when two nodes have similar roles but are not directly connected, for example because they lie on different branches of the control flow. 
To modulate this effect, we introduce a gating matrix $M \in \mathbb{R}^{|V| \times |V|}$, where each entry $M_{ij}$ controls how the absence score is explained. When $M_{ij}$ is large, edge absence is primarily attributed to role dissimilarity. When $M_{ij}$ is small, the fallback term $C$ becomes dominant, allowing the model to account for unconnected node pairs with similar roles.

\subsubsection{Learning Objective}
To train the model, our objective is to ensure that the learned latent roles are capable of reconstructing the original CFG structure as accurately as possible.
Formally, given the node attribute matrix $X$, we aim to maximize the marginal log-likelihood of the observed adjacency matrix $A$ under the generative model $p_{\theta}(\cdot)$:
\begin{equation}
\label{eq:obj_loglike}
    \log p_{\theta}(A|X)=\log \int p_{\theta}(A|Z)p(Z|X)dZ,
\end{equation}   
where $p_{\theta}(A \mid Z)$ models the likelihood of reconstructing the graph structure given $Z$, and $p(Z|X)=\prod_{i=1}^{|V|} p(\mathbf{z}_i | X)$ is the prior distribution of latent variables with $p(\mathbf{z}_i | X)=\text{LogitNormal}(\boldsymbol{\mu}_0,\Sigma_0)$.\footnote{We have implicitly assumed that $p(A|Z,X)=p(A|Z)$, that is, $A$ is conditionally independent of $X$ given $Z$, which is common in the literature \citep{kingma2014auto}. In practice, $(\boldsymbol{\mu}_0,\Sigma_0)$ can be set as $(\boldsymbol{0}, I)$.}

However, the integral over $Z$ in Equation~\eqref{eq:obj_loglike} is intractable, making direct optimization of the marginal log-likelihood computationally infeasible.
To address this, we derive the evidence lower bound (ELBO) as a tractable surrogate to the intractable marginal log-likelihood using variational inference and Jensen’s inequality~\citep{jensen1906fonctions, jordan1999introduction, kingma2014auto, rezende2014stochastic}:
\begin{equation}
\label{eq:elbo_details}
\log p_{\theta}(A|X) \geq \mathcal{L}_{\text{ELBO}}(\phi, \theta; G) =  \underbrace{\mathbb{E}_{q_{\phi}(Z|A, X)} \log {p_{\theta} (A|Z)}}_{\textbf{reconstruction loss}} - \underbrace{\text{KL} \big({q_{\phi} (Z|A, X)} \parallel p(Z|X) \big) }_{\textbf{KL divergence}}.
\end{equation}

The reconstruction loss term in Equation \eqref{eq:elbo_details} encourages the generative model to accurately reconstruct the adjacency matrix $A$ from the latent variables $Z$. Here, the expectation is taken over the variational distribution $q_{\phi}(Z|A, X)$, which is intended to approximate the true posterior $p_{\theta}(Z \mid A, X)$. In practice, the expectation is approximated by drawing a limited number of instances from $q_{\phi}(Z|A, X)$ based on the procedure defined by Equation \eqref{eq:z}.
The Kullback-Leibler (KL) divergence term in Equation \eqref{eq:elbo_details} regularizes the variational distribution $q_{\phi} (Z|A, X)$ to be close to the prior $p(Z|X)$.
Since both distributions factorize over nodes, this KL term decomposes as $\sum_{i=1}^{|V|}\text{KL}_i$, where each $\text{KL}_i$ is defined as:
{\small
\begin{equation}
\text{KL}_i = \text{KL}\big(q_{\phi}(\mathbf{z}_i|A,X)\,\|\,p(\mathbf{z}_i|X)\big)
= \frac{1}{2}\Big(
\text{tr}\big(\Sigma_0^{-1}\Sigma_i\big) 
+ (\boldsymbol{\mu}_0-\boldsymbol{\mu}_i)^{\top}\Sigma_0^{-1}(\boldsymbol{\mu}_0-\boldsymbol{\mu}_i) 
- K + \log \frac{|\Sigma_0|}{|\Sigma_i|} \Big).
\end{equation}
}

In summary, the learning objective is to maximize the ELBO with respect to both the variational parameters $\phi$ and the generative parameters $\theta$:
\begin{equation}
    \mathcal{L}_{\text{ELBO}}(\phi, \theta; G).
    \label{eq:elbo}
\end{equation}

\subsubsection{Operational Subgraph}
An \emph{operational subgraph} is a connected group of basic blocks in a CFG that are strongly associated with the same operational role. We extract operational subgraphs after estimating the latent roles of basic blocks, so that the identified roles can be explicitly incorporated into subsequent graph learning.
Specifically, for each node $v_i \in V$ in a CFG, we have the estimated latent role $\mathbf{z}_i \in \mathbb{R}^K$, where  $\mathbf{z}_{ik}$ denotes the probability that $v_i$ is associated with role $k$.\footnote{In practice, $\mathbf{z}_{ik}$ can be computed in accordance with Equation  \eqref{eq:z} by fixing the noise vector $\epsilon_i$ as a zero vector to obtain deterministic values for subsequent computations.}
To identify operational subgraphs, we first determine the set of nodes strongly affiliated with each role. For each role $k \in \{1, \dots, K\}$, we define a node set $V_k$ consisting of all nodes whose assignment probability to role $k$ exceeds a predefined threshold $\tau \in (0,1)$:
\begin{equation}
    V_k = \{v_i \in V \mid \mathbf{z}_{ik} > \tau \}.
\end{equation}

\begin{figure}[t]
\centering
\makebox[\textwidth][c]{
    \includegraphics[width=0.9\textwidth]{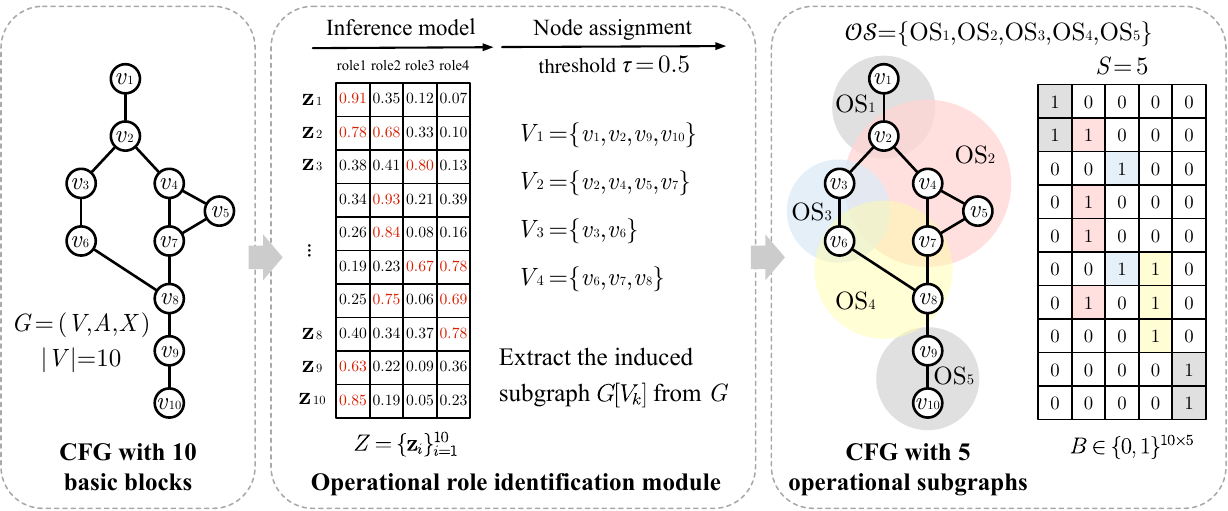}
}
\caption{Illustrative example of the operational role identification process.}
\label{fig:numeric_example}
\end{figure}

This thresholding operation allows a basic block to be associated with multiple roles while ensuring that only nodes with sufficiently high probability are assigned to that role.
Next, for each $V_k$, we extract the induced subgraph $G[V_k]$ from the original CFG by including all edges in $G$ that connect pairs of nodes within $V_k$. Since $G[V_k]$ may consist of multiple disconnected components, we further decompose it into its connected components.
Each connected component is then treated as an individual {operational subgraph} corresponding to role $k$, and the resulting collection is denoted as $\mathcal{O}\mathcal{S}_{k}$.
By repeating this procedure for all $K$ roles, we obtain a final set of operational subgraphs $\mathcal{O}\mathcal{S}$ from the CFG $G$:
\begin{equation}
    \mathcal{O}\mathcal{S}=\bigcup_{k=1}^{K} \mathcal{O}\mathcal{S}_{k} = \{ \text{OS}_1, \text{OS}_2, \cdots, \text{OS}_S\},
    \label{eq:sub}
\end{equation}
where $S$ is the total number of extracted operational subgraphs across all roles from CFG $G$.
We can represent the assignment of nodes to these operational subgraphs by a binary assignment matrix $B \in \{0,1\}^{|V| \times S}$, where $B_{is} = 1$ indicates that node $v_i$ in CFG $G$ belongs to the $s$-th operational subgraph $\text{OS}_s$, and $B_{is} = 0$ otherwise.

Figure~\ref{fig:numeric_example} illustrates a numeric example of the operational role identification process in MalGuard. 
Starting from a CFG $G$ with $|V|=10$ basic blocks, the inference model estimates the latent role variables $Z$ with $K=4$ latent roles. 
Based on a threshold of $\tau=0.5$, nodes are grouped into role-specific subsets, from which operational subgraphs ($S=5$) are derived according to Equation~\eqref{eq:sub}. 
The resulting binary assignment matrix $B \in \{0,1\}^{10 \times 5}$ encodes the membership relations between basic blocks and operational subgraphs, linking each node to its corresponding operational subgraph for subsequent graph learning.

\subsection{Program Graph Learning for Malware Detection}
\label{sec:4.2}

\begin{figure}[t]
\centering
\makebox[\textwidth][c]{
    \includegraphics[width=\textwidth]{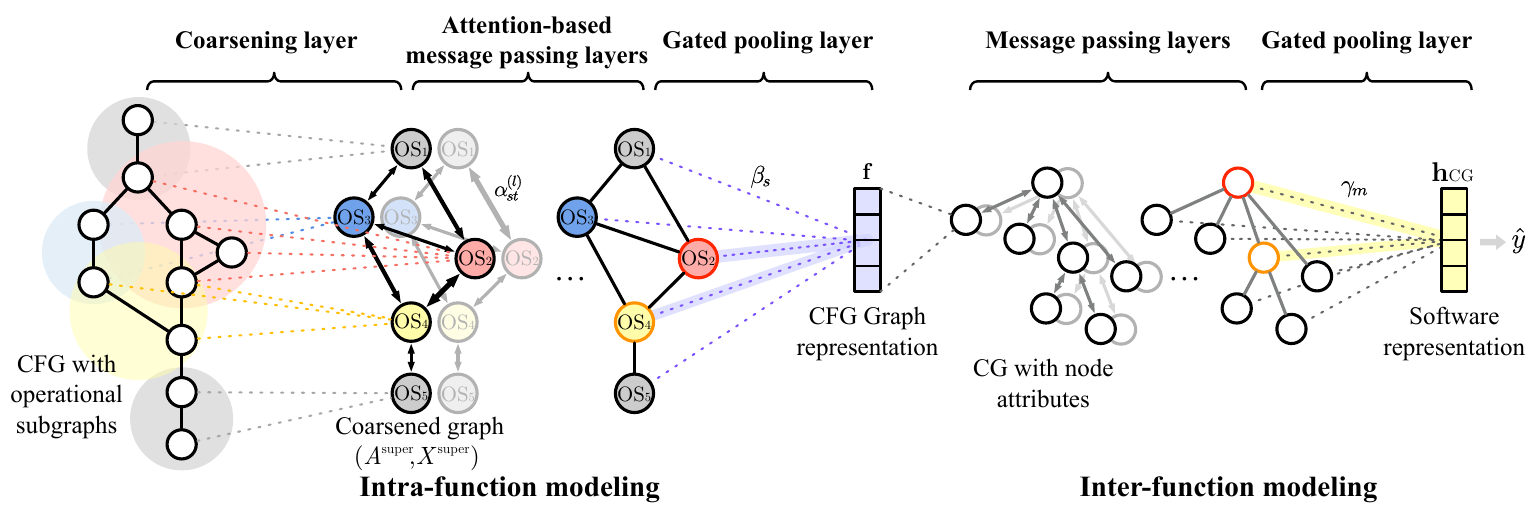}
}
\caption{The overall framework of the program graph representation learning method for malware detection.}
\label{fig:hier_gnn}
\end{figure}

This section introduces a program graph representation learning method for malware detection. It learns software representations by modeling interactions among operational roles, preserving sparse but critical malicious evidence, and capturing hierarchical program structure.  Figure~\ref{fig:hier_gnn} illustrates the overall framework.

Specifically, the method captures hierarchical program structure by modeling software at both the intra- and inter-function levels.
At the intra-function level, each function is represented as a CFG and encoded to model interactions among operational roles while preserving sparse but critical malicious signals within the function. The resulting function representations are then propagated to the CG. 
At the inter-function level, the CG is modeled to capture interactions among functions and to emphasize function-level signals that are most informative for malware detection, yielding the final software representation. We next describe the intra-function modeling, inter-function modeling, and learning objective.

\subsubsection{Intra-Function Modeling}
The goal of intra-function modeling is to learn a representation for each function by encoding its CFG. This module consists of three key layers: (i) a \emph{coarsening layer}, which converts the CFG from a graph of basic blocks into a graph of operational subgraphs, thereby enabling the model to capture interactions among operational roles;
(ii) a stack of \emph{attention-based message passing layers}, which aggregate information among operational subgraphs with learnable attention weights, allowing the model to preserve sparse but critical malicious signals from being diluted by routine neighbors during GNN aggregation; and (iii) a \emph{gated pooling layer}, which combines operational subgraph representations into a function representation through a learnable gate, allowing operational subgraphs carrying evidence relevant to malware detection to contribute more to the function representation.

Specifically, the coarsening layer first abstracts the CFG of basic blocks into a coarsened graph of operational subgraphs, which captures operational subgraphs and their complex interactions and facilitates later learning of meaningful program behavior.
Given a CFG $G=(V,A,X)$ and its assignment matrix $B \in \{0,1\}^{|V|\times S}$ that represents the assignment of $|V|$ nodes (in CFG $G$) to $S$ operational subgraphs, we construct the coarsened graph as follows. 
Each operational subgraph is treated as a \emph{super-node} that captures the collective role of its constituent basic blocks.
Edges between super-nodes encode execution dependencies between operational subgraphs. 
Formally, the coarsened graph is represented by a weighted adjacency matrix $A^{\text{super}} \in \mathbb{R}^{S \times S}$ and a super-node attribute matrix $X^{\text{super}} \in \mathbb{R}^{S \times d}$:
\begin{gather}
    A^{\text{super}} = B^\top A B, 
    \label{eq:super_adj} \\
    X^{\text{super}}_s= \sum_{v_i \in \text{OS}_s} \text{ReLU} \Big(\mathbf{W}_\text{super} \cdot \text{AGG} \left( X_{v_j}, \forall v_j \in \mathcal{N}_{\text{OS}_s}({v_i})\cup \{v_i\} \right) \Big),
    \label{eq:super_feat} 
\end{gather}
where $\text{AGG}$ is a permutation invariant aggregation function such as summation or average, $\mathcal{N}_{\text{OS}_s}(v_i)$ represents the set of neighboring basic blocks of basic block node $v_i$ in the subgraph $\text{OS}_s$, and $\mathbf{W}_\text{super}$ is a learnable weight matrix.
Equation~\eqref{eq:super_adj} combines the original adjacency matrix $A$ and the assignment matrix $B$ as a coarsened adjacency matrix $A^{\text{super}}$.
This operation compresses the original $|V|$ basic blocks into $S$ super-nodes, and aggregates edges between basic blocks belonging to different operational subgraphs into edges between their corresponding super-nodes. Each entry $A^{\text{super}}_{st}$ reflects the number of control flow edges between basic block nodes in $\text{OS}_s$ and $\text{OS}_t$, thereby encoding their execution dependencies.\footnote{This can be seen from the fact that $A^{\text{super}}_{st} = \sum_{i=1}^{|V|}\sum_{j=1}^{|V|} B_{is}\, A_{ij}\, B_{jt}
= \sum_{i \in \text{OS}_{s}}\sum_{j \in \text{OS}_{t}} A_{ij}$.}
Equation~\eqref{eq:super_feat} takes the operational subgraph $\text{OS}_s$ and the attributes of its constituent basic blocks as input, and applies a single-layer GNN specified by $\mathbf{W}_\text{super}$ to generate the representation of this subgraph.
The resulting representation of $\text{OS}_s$ serves as the attribute vector $X^{\text{super}}_s$ for super-node $v^{\text{super}}_s$, providing a compact summary of the subgraph semantics. 

With the coarsened graph $(A^{\text{super}}, X^{\text{super}})$, the next step is to update each operational subgraph representation by aggregating information from its neighbors.
Message passing GNNs are a natural choice here, owing to their expressive power to model graph-structured data~\citep{kipf2017semisupervised, hamilton2017inductive}. A common practice in standard message passing is to uniformly aggregate information from all neighbors when updating a node representation. However, this practice is not ideal for malware detection and can even undermine detection accuracy. 
In practice, malware typically contains only a small fraction of operational subgraphs that carry malicious cues, while the majority reflect common behaviors~\citep{alam2015annotated}. Consider the case where a malicious operational subgraph is adjacent to common ones; its signal would be overwhelmed during aggregation with common neighbors, causing the model to overlook these sparse yet critical malicious patterns.
This calls for a more selective message passing mechanism that can adaptively assign different importance to neighbors during aggregation. 

To this end, we introduce an attention-based message passing layer to model the relative importance of different neighboring super-nodes during message passing. This design is inspired by the graph attention mechanism~\citep{velickovic2018graph}, where learnable attention weights determine the contribution of each neighbor when updating a super-node representation.
Specifically, the representation of super-node $v^{\text{super}}_s$ is updated by aggregating information from its neighboring super-nodes $v^{\text{super}}_t \in \mathcal{N}(v^{\text{super}}_s)$, with each neighbor’s influence reflected by the learnable attention weight $\alpha^{(l)}_{st}$.
This process can be formulated as:
\begin{equation}
\mathbf{g}_{s}^{(l)} = \text{ReLU} \left(\mathbf{W}_\text{role}^{(l)} \cdot \text{AGG} \bigl( \alpha^{(l)}_{st}\mathbf{g}_{t}^{(l-1)}, \forall v^{\text{super}}_t \in \mathcal{N}(v^{\text{super}}_s)\right),
\label{eq:role}
\end{equation}
where $\mathbf{g}_s^{(0)} = X^{\text{super}}_s$, $\mathbf{W}_\text{role}^{(l)} \in \mathbb{R}^{d \times d}$ is a learnable weight matrix at layer $l$, and $\mathcal{N}(v^{\text{super}}_s) = \{ v^{\text{super}}_t \mid A^{\text{super}}_{st} > 0 \} \cup \{v^{\text{super}}_s \}$ denotes the neighboring super-nodes of $v^{\text{super}}_s$ in the coarsened graph.

The attention weight $\alpha_{st}^{(l)}$ is learned from the representations of the focal super-node $v^{\text{super}}_s$ and its neighbor $v^{\text{super}}_t$ (i.e., $\mathbf{g}_{s}^{(l-1)}$ and $\mathbf{g}_{t}^{(l-1)}$), thereby reflecting the relative influence of neighbor $v^{\text{super}}_t$ on $v^{\text{super}}_s$.
For example, one possible outcome is that when $v^{\text{super}}_s$ carries malicious signal while $v^{\text{super}}_t$ reflects common behaviors, the learned $\alpha_{st}^{(l)}$ becomes lower, preventing the malicious signal in $v^{\text{super}}_s$ from being overwhelmed by common behaviors.
Moreover, we incorporate the edge weights $A^{\text{super}}_{st}$ into the attention computation, ensuring that execution dependencies observed more frequently exert greater influence during message passing.
Formally, the attention weight is defined as:
\begin{equation}
\alpha^{(l)}_{st} = \frac{
A^{\text{super}}_{st} \ \exp\left( \text{LeakyReLU}\left( \mathbf{a}^\top \left[ \mathbf{W}_\text{attn} \mathbf{g}_{s}^{(l-1)} \Big\Vert \mathbf{W}_\text{attn} \mathbf{g}_{t}^{(l-1)} \right] \right) \right)
}{
\sum_{v^{\text{super}}_r \in \mathcal{N}(v^{\text{super}}_s)} A^{\text{super}}_{sr} \ \exp\left( \text{LeakyReLU}\left( \mathbf{a}^\top \left[ \mathbf{W}_\text{attn} \mathbf{g}_{s}^{(l-1)} \Big\Vert \mathbf{W}_\text{attn} \mathbf{g}_{r}^{(l-1)} \right] \right) \right)
}.
\label{eq:attn}
\end{equation}

In this equation, each super-node representation is first linearly transformed by a shared weight matrix $\mathbf{W}_\text{attn} \in \mathbb{R}^{d \times d}$. The transformed representations are then concatenated (denoted by $\Vert$) and projected onto a scalar score by a learnable attention vector $\mathbf{a} \in \mathbb{R}^{2d}$. 
This score is weighted by the edge weight $A^{\text{super}}_{st}$ and passed through a LeakyReLU activation. 
Finally, the normalization ensures that $\sum_{v^{\text{super}}_t \in \mathcal{N}(v^{\text{super}}_s)} \alpha^{(l)}_{st} = 1$, making the attention weights comparable across the neighbors of $v^{\text{super}}_s$.
By stacking $L$ layers of attention-based message passing, we obtain the final representation $\mathbf{g}_{s}^{(L)}$ for each operational subgraph $\text{OS}_s \in \mathcal{O}\mathcal{S}$.

Finally, the gated pooling layer is designed to combine the operational subgraph representations into a unified CFG representation through a learnable fusion gate.
Unlike traditional mean or sum pooling, which treats all operational subgraph representations equally and may therefore dilute the contribution of rare but decisive subgraphs for malware detection, the \emph{fusion gate}  learns how much each operational subgraph should contribute to the final CFG representation.
Specifically, the fusion gate
$\boldsymbol{\beta} = (\beta_1,\beta_2,\ldots,\beta_S)^\top \in \mathbb{R}^{S}$
is computed from the representations of operational subgraphs and serves as a set of learnable weights that suppress redundant or noisy signals while amplifying rare but decisive ones. This design ensures that the final CFG representation $\mathbf{f}\in\mathbb{R}^d$ preserves the most informative semantics for malware detection. Formally, the process can be expressed as:
\begin{gather}
    \mathbf{f} = \sum_{\text{OS}_s \in \mathcal{O}\mathcal{S}}
    \beta_s \mathbf{g}_s^{(L)}, \quad
    \beta_s = 
    \frac{\exp\big( \mathbf{W}_{\text{CFG}}^\top \mathbf{g}_s^{(L)} \big)}
    {\sum_{\text{OS}_r \in \mathcal{O}\mathcal{S}} \exp\big( \mathbf{W}_{\text{CFG}}^\top \mathbf{g}_r^{(L)} \big)},
    \label{eq:CFGgate}
\end{gather}
where $\mathbf{W}_{\text{CFG}} \in \mathbb{R}^d$ is a learnable vector.
The CFG representations obtained from all functions are then stacked to form $F_{\text{CFG}}\in\mathbb{R}^{|V_{\text{CG}}|\times d}$.

\subsubsection{Inter-Function Modeling}
Given the CG with node attribute matrix $F_{\text{CFG}}$, we apply a stack of message passing layers to model how functions interact through the call structure and collectively realize global software behaviors.
At the $l$-th layer of the GNN, the representation of a function node $v_m$ is updated by aggregating information from its neighbors and itself through an aggregation function $\text{AGG}$, followed by a linear transformation and a nonlinear activation:
\begin{equation}
     \mathbf{h}_{m}^{(l)} = \text{ReLU} \left( \mathbf{W}_{\text{func}}^{(l)} \cdot \text{AGG} \left( \mathbf{h}_{n}^{(l-1)}, \forall {v_n} \in \mathcal{N}({v_m})\cup \{v_m\} \right) \right),
    \label{eq:func}
\end{equation}
where $ \mathbf{h}_{m}^{(l)}$ represents the representation of function node $v_m$ at layer $l$, $\mathbf{h}_m^{(0)}$ is the $m$-th row of $F_{\text{CFG}}$, $\mathcal{N}({v_m})$ is the set of neighboring function nodes of $v_m$ in the CG, and $\mathbf{W}_{\text{func}}^{(l)} \in \mathbb{R}^{{d} \times {d}}$ is a learnable weight matrix. 
By stacking $L$ message passing layers, we obtain the final function representations $\mathbf{h}_{m}^{(L)}$ for $v_m \in V_{\text{CG}}$.

Then, we adopt a gated pooling layer similar to the one used when performing intra-function modeling.
This layer learns importance scores over function nodes, allowing functions carrying signals relevant to malware detection to contribute more to the final software representation.
Specifically, the fusion gate $\boldsymbol{\gamma}=(\gamma_1,\gamma_2,\ldots,\gamma_{|V_{\text{CG}}|})^\top \in \mathbb{R}^{|V_{\text{CG}}|}$
assigns an importance weight to each function node. 
The software representation $\mathbf{h}_{\text{CG}}$ is then computed as:
\begin{gather}
    \mathbf{h}_{\text{CG}} = \sum_{v_m \in V_{\text{CG}}}
    \gamma_m \mathbf{h}_{m}^{(L)}, \quad
    \gamma_m = 
    \frac{\exp\big( \mathbf{W}_{\text{CG}}^\top \mathbf{h}_{m}^{(L)} \big)}
    {\sum_{v_n \in V_{\text{CG}}} \exp\big( \mathbf{W}_{\text{CG}}^\top \mathbf{h}_{n}^{(L)} \big)},
    \label{eq:CGgate}
\end{gather}
where $\gamma_m$ is the learned importance weight assigned to function node $v_m$, and $\mathbf{W}_{\text{CG}} \in \mathbb{R}^d$ is a learnable vector.

\subsubsection{Learning Objective}

Finally, the software representation $\mathbf{h}_{\text{CG}}$ is passed through an MLP layer parameterized by $\psi$, followed by a sigmoid activation, to produce the malware detection probability of the input software:
\begin{equation}
    \hat{y} = \text{sigmoid}\left( \text{MLP}_{\psi}( \mathbf{h}_{\text{CG}} ) \right).
    \label{eq:y}
\end{equation}
A higher value of $\hat{y}$ indicates a greater likelihood that the software is malicious.

The malware detection task can be formulated as a binary classification problem, where the objective is to determine whether a given software is malicious or benign. To this end, the framework is trained by minimizing the binary cross-entropy loss between the predicted probability $\hat{y}$ and the ground-truth label $y \in \{0,1\}$:
\begin{equation}
\mathcal{L}_{\text{BCE}}(\Theta) = - \left( y \log \hat{y} + (1 - y) \log (1 - \hat{y}) \right),
\label{eq:bce}
\end{equation}
where  $\Theta = \{\mathbf{W}_\text{super}, \mathbf{W}_{\text{role}}, \mathbf{W}_{\text{attn}}, \mathbf{W}_{\text{func}}, \mathbf{W}_{\text{CFG}}, \mathbf{W}_{\text{CG}}, \mathbf{a}, \psi\}$
are the trainable parameters.

\subsection{Overall Algorithm}
\label{sec:4.3}
We integrate the components introduced above into a two-stage training procedure. 
The first stage performs {operational role identification}, which infers latent roles for basic blocks and extracts operational subgraphs. 
The second stage performs {program graph learning}, which learns software representations from  hierarchical program graphs for malware detection.

\begin{algorithm}[t]
\caption{The Overall Training Procedure of MalGuard}
\label{alg:MalGuard}
\small
\linespread{0.9}\selectfont
\textbf{Input:} A labeled set of software $\mathcal{R}$ \\
\textbf{Output:} Learned model parameters $\phi, \theta, \mathbf{W}_\text{super},\mathbf{W}_{\text{role}}, \mathbf{W}_{\text{attn}}, \mathbf{W}_{\text{func}}, \mathbf{W}_{\text{CFG}}, \mathbf{W}_{\text{CG}}, \mathbf{a}, \psi$

\begin{algorithmic}[1]
\State Construct the hierarchical graph $\mathcal{H} = (G_{\text{CG}}, \mathcal{G}_{\text{CFG}})$ for each software instance in $\mathcal{R}$, let $\mathcal{G}_{\text{CFG}}^{\text{all}}$ denote the collection of all CFGs from these hierarchical program graphs
\label{line:1}

\Statex \textit{// Stage 1: Operational Role Identification}

\For{$\text{epoch}_1 \in 1:N_{\text{epoch}_1}$}
\label{line:2}
\Comment{$N_{\text{epoch}_1}$: number of epochs in Stage 1}

    \For{each mini-batch of CFGs sampled from $\mathcal{G}_{\text{CFG}}^{\text{all}}$}    
    \label{line:3}

        \State Compute $\boldsymbol{\mu}, \boldsymbol{\sigma}$ for each CFG in the mini-batch
        \label{line:4}
        \Comment{Equation~\eqref{eq:mu}}
        
        \State Compute $Z$ for each CFG in the mini-batch
        \label{line:5}
        \Comment{Equation~\eqref{eq:z}}
        
        \State Compute batch loss $-\mathcal{L}_{\text{ELBO}}$ over the mini-batch
        \label{line:6}
        \Comment{Equation~\eqref{eq:elbo}}
        
        \State Update $\phi$ and $\theta$ via gradient descent on the batch loss
        \label{line:7}
        
    \EndFor
    \label{line:8}

\EndFor
\label{line:9}

\For{each CFG $G=(V,A,X)\in\mathcal{G}_{\text{all}}$}
\label{line:10}    
    \State Extract operational subgraphs $\mathcal{O}\mathcal{S} = \{\text{OS}_1, \dots, \text{OS}_S\}$
    \label{line:11}
    
    \State Construct $B \in \{0,1\}^{|V| \times S}$
    \label{line:12}
\EndFor
\label{line:13}

\Statex \textit{// Stage 2: Program Graph Learning for Malware Detection}
\For{$\text{epoch}_2 \in 1:N_{\text{epoch}_2}$}
\label{line:14}
\Comment{$N_{\text{epoch}_2}$: number of epochs in Stage 2}

 \For{each software instance in $\mathcal{R}$}
 \label{line:15}
    
    \State Retrieve its hierarchical program graph $\mathcal{H}=(G_{\text{CG}},\mathcal{G}_{\text{CFG}})$
    \label{line:16}

    \For{each CFG $G=(V,A,X)\in\mathcal{G}_{\text{CFG}}$}
    \label{line:17}
    
        \State Retrieve $\mathcal{O}\mathcal{S}$ and $B$ associated with CFG $G$
        \label{line:18}
        
        \State Compute $A^{\text{super}} = B^\top A B$ and $X^{\text{super}}$
        \label{line:19}
        \Comment{Equation~\eqref{eq:super_adj},~\eqref{eq:super_feat}}
        
        \State Compute $\alpha^{(l)}_{st}$, $\mathbf{g}_s^{(l)}$ for $l \in 1:L$
        \label{line:20}
        \Comment{Equation~\eqref{eq:role},~\eqref{eq:attn}}
        
        \State Compute and store the CFG representation $\mathbf{f}$
        \label{line:21}
        \Comment{Equation~\eqref{eq:CFGgate}}
    \EndFor
    \label{line:22}
    
    \State Stack stored CFG representations as $F_{\text{CFG}}$ for $G_{\text{CG}}$
    \label{line:23}
    
    \State Compute $\mathbf{h}_{\text{CG}}$ 
    \label{line:24}
    \Comment{Equation~\eqref{eq:func},~\eqref{eq:CGgate}}

    \State Compute $\hat{y} = \text{sigmoid}\left( \text{MLP}_{\psi}( \mathbf{h}_{\text{CG}} ) \right)$
    \label{line:25}    
    \Comment{Equation~\eqref{eq:y}}
    
    \State Compute loss $\mathcal{L}_{\text{BCE}}$
    \label{line:26}
    \Comment{Equation~\eqref{eq:bce}}
    
    \State Update $\mathbf{W}_\text{super}, \mathbf{W}_{\text{role}}, \mathbf{W}_{\text{attn}}, \mathbf{W}_{\text{func}}, \mathbf{W}_{\text{CFG}}, \mathbf{W}_{\text{CG}}, \mathbf{a}, \psi$ with gradient descent on $\mathcal{L}_{\text{BCE}}$
    \label{line:27}
\EndFor
\label{line:28}
\EndFor
\label{line:29}

\State \Return $\phi, \theta, \mathbf{W}_\text{super}, \mathbf{W}_{\text{role}}, \mathbf{W}_{\text{attn}}, \mathbf{W}_{\text{func}}, \mathbf{W}_{\text{CFG}}, \mathbf{W}_{\text{CG}}, \mathbf{a}, \psi$
\label{line:30}
\end{algorithmic}
\end{algorithm}

The overall training procedure of MalGuard is summarized in Algorithm~\ref{alg:MalGuard}.
Stage~1, operational role identification, corresponds to lines~\ref{line:2}--\ref{line:13}.
In this stage, MalGuard trains the latent role identification module over mini-batches of CFGs sampled from $\mathcal{G}^{\text{all}}_{\text{CFG}}$.
For each mini-batch, it computes the parameters of the latent role distribution (line~\ref{line:4}), infers latent role variables $Z$ for each CFG (line~\ref{line:5}), evaluates the batch objective $-\mathcal{L}_{\text{ELBO}}$ (line~\ref{line:6}), and updates $\phi$ and $\theta$ by gradient descent (line~\ref{line:7}). After training, the inferred role affiliations are used to extract operational subgraphs and construct the assignment matrix $B$ for each CFG (lines~\ref{line:10}--\ref{line:13}).
{Stage 2, program graph learning for malware detection, corresponds to lines~\ref{line:14}--\ref{line:29}.
In this stage, MalGuard learns a representation of the hierarchical program graph for each software instance for malware detection.
The intra-function modeling includes the coarsening layer, attention-based message passing layers, and gated pooling layer, which together produce a CFG-level representation for each function (lines~\ref{line:19}--\ref{line:21}). The inter-function modeling then applies message passing layers and a gated pooling layer over the CG to obtain the software-level representation $\mathbf{h}_{\text{CG}}$ (line~\ref{line:24}).
Finally, MalGuard computes the malware prediction $\hat{y}$ (line~\ref{line:25}), evaluates the binary cross-entropy loss $\mathcal{L}_{\mathrm{BCE}}$ (line~\ref{line:26}), and updates the program graph learning parameters by gradient descent (line~\ref{line:27}).}
The computational complexity analysis of MalGuard can be found in Appendix~\ref{appendix:complexity}.

\section{Experiments}
This section provides a comprehensive evaluation of the proposed method on a real-world, manually collected software dataset.
We first describe the experimental setup in Section~\ref{sec:5.1}. 
Then, we evaluate the malware detection performance of MalGuard against various benchmarks in Section~\ref{sec:5.2} and conduct ablation studies to evaluate the contribution of key components in MalGuard in Section~\ref{sec:5.3}.
Section~\ref{sec:5.4} provides a case study to demonstrate the interpretability of our model, and
Section~\ref{sec:5.5} presents an economic value analysis.

\subsection{Experimental Setup}
\label{sec:5.1}
\vpara{Dataset.}
We manually constructed a dataset of malware and benign software from diverse sources, ensuring that it is both up-to-date and categorically diverse. 
For malware, we collected all newly released instances between February 14, 2024, and October 4, 2024 from VirusShare.com\footnote{\href{https://virusshare.com/}{\texttt{https://virusshare.com/}}}, a widely used repository that is continuously updated with new malware from multiple sources and provides researchers with access to live malware. 
These malware instances span a wide range of families, including trojans, ransomware, worms, and others, thereby reflecting the diversity of real-world malware.
For benign software, we sourced applications from several reputable freeware websites, including Wiki DLL\footnote{\href{https://wikidll.com/}{\texttt{https://wikidll.com/}}}, NirSoft\footnote{\href{https://www.nirsoft.net/}{\texttt{https://www.nirsoft.net/}}}, and CNET\footnote{\href{https://download.cnet.com/}{\texttt{https://download.cnet.com/}}}, which collectively offer diverse benign software across different domains. 
To ensure the reliability of ground truth labels,  all instances were cross-verified with VirusTotal\footnote{\href{https://virustotal.com/}{\texttt{https://virustotal.com/}}}, 
an industry-standard platform that provides reports of malware detection outcomes from over 70 commercial antivirus engines.
Each collected instance was compared against VirusTotal reports, and instances showing inconsistencies with their ground-truth labels were excluded to reduce potential mislabeling. In total, 24 such instances were identified and removed from the dataset.
To further reflect real-world class imbalance, where benign software is typically more prevalent, we adopted a 1:5 ratio of malicious to benign instances.
This results in the final dataset containing 1,428 malware instances and 7,342 benign instances.

\vpara{Benchmark methods.}
We benchmarked our method against representative ML malware detection methods.
First, we compared our method with byte-based methods, which operate directly on the raw byte sequences of the program.
The comparison with these methods demonstrates the advantage of graph-based methods in defending against malware with evasive behaviors.
\begin{itemize}
    \item MalConv~\citep{raff2018malware}, which is a prominent state-of-the-art benchmark that embeds byte sequences into a vector space and applies a gated convolutional network for detection.
    \item NonNeg~\citep{fleshman2018non}, which is a variant of MalConv designed to detect evasive malware by constraining the final layer weights to be non-negative.
    \item DRSM~\citep{saha2024drsm}, another MalConv variant that improves robustness against evasive behaviors by dividing input bytes into segments, applying a pre-trained MalConv model to each segment, and combining the results through a voting strategy.
\end{itemize}

We further evaluated our method against graph-based ML malware detection methods that detect malware by representing software as program graphs to capture execution logic and mitigate evasive behaviors. This comparison demonstrates the added value of our method in modeling program behaviors more effectively, thereby enabling more accurate malware detection.
\begin{itemize}
    \item GCN~\citep{kipf2017semisupervised}, GAT~\citep{velickovic2018graph}, GIN~\citep{xu2018how}, which are standard GNN baselines applied to hierarchical program graphs for malware detection.
    \item MAGIC~\citep{yan2019classifying}, which leverages programs’ CFGs and applies DGCNN~\citep{zhang2018end} to these graphs for malware detection.
    \item DeepCall~\citep{chen2023deepcall}, which builds on programs’ CGs and also employs DGCNN for malware detection.
    \item Mal2GCN~\citep{kargarnovin2024mal2gcn}, which leverages programs’ CGs and incorporates a non-negative weight constraint on GNN to enhance robustness against evasive malware.
    \item MalGraph~\citep{ling2022malgraph}, which exploits hierarchical program graphs and introduces hierarchical graph learning for detection.
\end{itemize}

\vpara{Evaluation metrics.}
Considering the inherent class imbalance in malware detection, evaluation requires metrics that go beyond overall accuracy. We therefore adopted three evaluation metrics: F1-score, AUC (the area under the receiver operating characteristic curve, or ROC curve), and AUPRC (the area under the precision--recall curve).
In all cases, malware instances were treated as the positive class and benign software as the negative class. Specifically, the F1-score is the harmonic mean of Precision and Recall, where Recall measures the proportion of malware correctly identified and Precision measures the proportion of predicted malware that are truly malicious, i.e., $\text{F1} = 2 (\text{Precision} \cdot \text{Recall}) / (\text{Precision} + \text{Recall})$. 
AUC is computed as the area under the ROC curve, which plots the true positive rate against the false positive rate at varying thresholds. It reflects the probability that a randomly chosen malware instance receives a higher detection score than a benign one. AUPRC is computed as the area under the Precision-Recall curve, which plots precision against recall at varying thresholds and is particularly informative under class imbalance.
The dataset was split into training, validation, and testing sets using an 8:1:1 ratio for model evaluation.

\vpara{Implementation details.}
We conducted all experiments on a single NVIDIA GeForce RTX 3090 GPU with 24GB memory. All models were implemented in Python 3.8 with PyTorch library.
Specifically, when implementing our method, we set the hidden dimension $d$ to 128, the number of GNN layers $L$ to 2, all MLPs were composed of 2 layers, and all aggregation functions $\text{AGG}(\cdot)$ were implemented as the mean operator. In the operational role identification approach, the fallback likelihood for edge absence $C$ was fixed at 0.3, the role assignment threshold $\tau$ was set to 0.5, and we adopted a standard multivariate logit-normal prior with $\boldsymbol{\mu}_0 = \mathbf{0}$ and $\Sigma_0 = I$. 
For optimization, we employed the Adam optimizer~\citep{kingma2015adam} with a learning rate of $1 \times 10^{-3}$ and a batch size of 64. Training was performed with early stopping based on the validation loss to mitigate overfitting. 
To construct program graphs, we disassembled all software instances in the dataset using the Python program analysis framework angr\footnote{\href{https://angr.io/}{\texttt{https://angr.io/}}}, extracting CGs and CFGs.
The original node attributes of CFG were initialized with 13 statistical features, including the numbers of call, transfer, arithmetic, logic, compare, move, and termination instructions, as well as counts of data declarations, total instructions, string constants, integer constants, in-degree, and out-degree.

For fair comparisons, the hyperparameters of all baseline models were tuned to their optimal settings. MalConv, NonNeg, MAGIC, DeepCall, and Mal2GCN were strictly re-implemented following their original papers.
The GNN-based baselines (GCN, GAT, and GIN) were implemented using the PyG library. 
For DRSM and MalGraph, we utilized the official implementations released by the authors.
Details can be found in Appendix~\ref{appendix:experiment}.

\subsection{Malware Detection Performance}
\label{sec:5.2}

\begin{table}[t]
    \centering
    \fontsize{9}{9}\selectfont
    \caption{Malware detection performance of our method compared with byte-based and graph-based benchmarks. The best and second-best performances are highlighted in \textbf{bold} and \underline{underline}, respectively. All values are reported as means with standard deviations in parentheses. Relative improvements of MalGuard over benchmarks are reported.}
    \label{tab:main}
    \begin{tabular}{ll|cc|cc|cc}
        \toprule
        Category & Method & F1-score & \makecell{Improvement by \\ MalGuard}&  AUC & \makecell{Improvement by \\ MalGuard}  & AUPRC & \makecell{Improvement by \\ MalGuard} \\
        \midrule
        \multirow{5}{*}{Byte-Based}
            & MalConv  
            & \makecell{87.03$^{***}$\\(0.53)} & 6.06\% 
            & \makecell{89.27$^{***}$\\(0.40)} & 5.96\% 
            & \makecell{\underline{89.67}$^{**}$\\(0.48)} & 3.91\% \\
            & NonNeg  
            & \makecell{80.50$^{***}$\\(2.35)} & 14.66\% 
            & \makecell{84.16$^{***}$\\(1.68)} & 12.39\% 
            & \makecell{85.66$^{**}$\\(1.64)} & 8.78\% \\
            & DRSM 
            & \makecell{87.21$^{**}$\\(1.15)} & 5.84\% 
            & \makecell{90.06$^{**}$\\(1.38)} & 5.03\% 
            & \makecell{89.37$^{***}$\\(0.59)} & 4.26\% \\
        \midrule
        \multirow{13}{*}{Graph-Based}
            & GCN     
            & \makecell{82.90$^{**}$\\(2.60)} & 11.34\% 
            & \makecell{87.89$^{*}$\\(2.80)} & 7.62\% 
            & \makecell{85.50$^{***}$\\(1.52)} & 8.98\% \\
            & GAT     
            & \makecell{84.47$^{***}$\\(1.51)} & 9.27\% 
            & \makecell{90.75$^{*}$\\(1.16)} & 4.23\% 
            & \makecell{85.76$^{***}$\\(1.55)} & 8.65\% \\
            & GIN     
            & \makecell{81.25$^{***}$\\(2.18)} & 13.60\% 
            & \makecell{87.67$^{**}$\\(2.60)} & 7.89\% 
            & \makecell{83.52$^{***}$\\(1.11)} & 11.57\% \\
            & MAGIC   
            & \makecell{83.47$^{***}$\\(2.44)} & 10.58\% 
            & \makecell{89.01$^{**}$\\(1.80)} & 6.27\% 
            & \makecell{85.25$^{***}$\\(2.06)} & 9.30\% \\
            & DeepCall
            & \makecell{\underline{87.77}$^{***}$\\(1.72)} & 5.16\% 
            & \makecell{\underline{91.44}$^{***}$\\(0.87)} & 3.44\% 
            & \makecell{89.21$^{**}$\\(1.69)} & 4.45\% \\
            & Mal2GCN 
            & \makecell{77.33$^{***}$\\(1.60)} & 19.36\% 
            & \makecell{81.86$^{***}$\\(0.99)} & 15.55\% 
            & \makecell{82.87$^{***}$\\(1.33)} & 12.44\% \\
            & MalGraph
            & \makecell{82.76$^{**}$\\(3.47)} & 11.53\% 
            & \makecell{87.49$^{*}$\\(3.68)} & 8.12\% 
            & \makecell{85.78$^{**}$\\(2.18)} & 8.63\% \\
        \midrule
        \multicolumn{2}{c|}{MalGuard}
            & \makecell{\textbf{92.30}\\(0.73)} & -- 
            & \makecell{\textbf{94.59}\\(0.70)} & -- 
            & \makecell{\textbf{93.18}\\(0.63)} & -- \\
        \bottomrule
    \end{tabular}
    \par\smallskip
    {\textit{Note.} Statistical significance is marked as * ($p<0.05$), ** ($p<0.01$), and *** ($p<0.001$).}
\end{table}

We first evaluated the effectiveness of MalGuard on malware detection. Table~\ref{tab:main} reports the results on our collected dataset. Notably, MalGuard consistently achieves the best performance, with relative improvements of 5.16\% in F1-score, 3.44\% in AUC, and 3.91\% in AUPRC compared with the best-performing benchmark.
All improvements achieved by MalGuard are statistically significant ($p<0.05$).
Our method’s superiority over byte-based methods arises from its ability to model programs as program graphs, which effectively capture execution logic and provide inherent robustness against evasive malware.
The performance gap between our method and other graph-based methods underscores the importance of explicitly identifying and incorporating operational roles, while jointly modeling hierarchical program structure and preserving sparse malicious signals.
More specifically, existing graph-based benchmarks commonly overlook operational roles and fail to adequately capture malicious signals.
GCN, GIN, and GAT serve as standard GNN baselines that are not tailored for malware detection. 
MAGIC, DeepCall and Mal2GCN are specifically designed for malware detection, but they model programs either as CFGs or CGs and thus cannot exploit hierarchical program structure.
MalGraph models programs as hierarchical graphs, but it fails to disentangle operational roles and overlooks malicious signals, resulting in weaker performance.

We further evaluated MalGuard and the benchmark methods under more challenging conditions by injecting evasive behaviors into the malware instances in our test set with six representative evasion techniques. Results indicate that MalGuard achieves improvements over all benchmarks, particularly byte-based methods, underscoring its strong capability to defend against evasive malware. Detailed results are provided in Appendix~\ref{appendix:evasive}.

\subsection{Ablation Studies}
\label{sec:5.3}
We conducted ablation studies to evaluate the contribution of key components in MalGuard. Specifically, we assessed five variants of our method.
(1) MalGuard-OR removes the operational role identification module in our method, operating without both the role identification and the role-based coarsening layer. 
(2) MalGuard-Hier overlooks the hierarchical program structure and learns only from CFGs, producing the final software representation by directly pooling CFG representations without inter-function modeling.
(3) MalGuard-CL removes the coarsening layer and instead incorporates the role information as an additional node attribute in CFGs.
(4) MalGuard-Attn removes the attention-based message passing layers and instead adopts the standard message passing as GCN~\citep{kipf2017semisupervised}.
(5) MalGuard-GP removes the gated pooling layer and instead uses sum pooling as GIN~\citep{xu2018how}.

\begin{table}[t]
    \centering
    \fontsize{10}{10}\selectfont
    \caption{Ablation study results. $\Delta$ values show the relative performance drop observed after removing each component from MalGuard.}
    \label{tab:ablation}
    \begin{tabular}{lccc}
        \toprule
        Method & F1-score ($\Delta$) & AUC ($\Delta$) & AUPRC ($\Delta$) \\
        \midrule
        MalGuard
        & \textbf{92.30}
        & \textbf{94.59}
        & \textbf{93.18} \\ 
        \midrule
        MalGuard-OR
        & 86.69 ($\downarrow$6.08\%) 
        & 89.12 ($\downarrow$5.78\%) 
        & 87.39 ($\downarrow$6.21\%) \\            
        MalGuard-Hier
        & 86.95 ($\downarrow$5.80\%) 
        & 90.31 ($\downarrow$4.52\%) 
        & 88.06 ($\downarrow$5.49\%) \\
        MalGuard-CL
        & 89.78 ($\downarrow$2.73\%) 
        & 92.57 ($\downarrow$2.14\%) 
        & 90.40 ($\downarrow$2.98\%) \\            
        MalGuard-Attn
        & 90.77 ($\downarrow$1.66\%) 
        & 92.80 ($\downarrow$1.89\%) 
        & 91.65 ($\downarrow$1.64\%) \\           
        MalGuard-GP
        & 90.56 ($\downarrow$1.89\%) 
        & 92.78 ($\downarrow$1.91\%) 
        & 91.34 ($\downarrow$1.97\%) \\ 
        \bottomrule
    \end{tabular}
\end{table}

The results are summarized in Table~\ref{tab:ablation}. We observe that all variants perform worse than MalGuard, underscoring the effectiveness of each component. Notably, removing the operational role identification (MalGuard-OR) results in the most severe degradation, highlighting the necessity of identifying and leveraging operational roles in malware detection. The performance decrease of MalGuard-Hier demonstrates the importance of jointly modeling intra- and inter-function levels in the hierarchical program graph. Moreover, the performance drops observed for MalGuard-CL, MalGuard-Attn, and MalGuard-GP indicate the effectiveness of the coarsening layer, attention-based message passing layer, and gated pooling layer in modeling interactions among operational roles, preserving malicious evidence while modeling the hierarchical graph structure.

\subsection{Case Study}
\label{sec:5.4}
To demonstrate how MalGuard identifies semantically meaningful operational roles and provides interpretable insights into malware behavior, we conduct a case study on a representative malware instance, ``MemoryDiagnostic.exe''. Specifically, our objective is to examine whether the operational roles uncovered by MalGuard correspond to meaningful behavioral patterns that align with expert analysis, and whether such abstractions can reveal latent malicious intent that is not observable at the level of individual basic blocks.

\begin{figure}[t]
\centering
\vspace{-0.2in}
\includegraphics[width=0.75\textwidth]{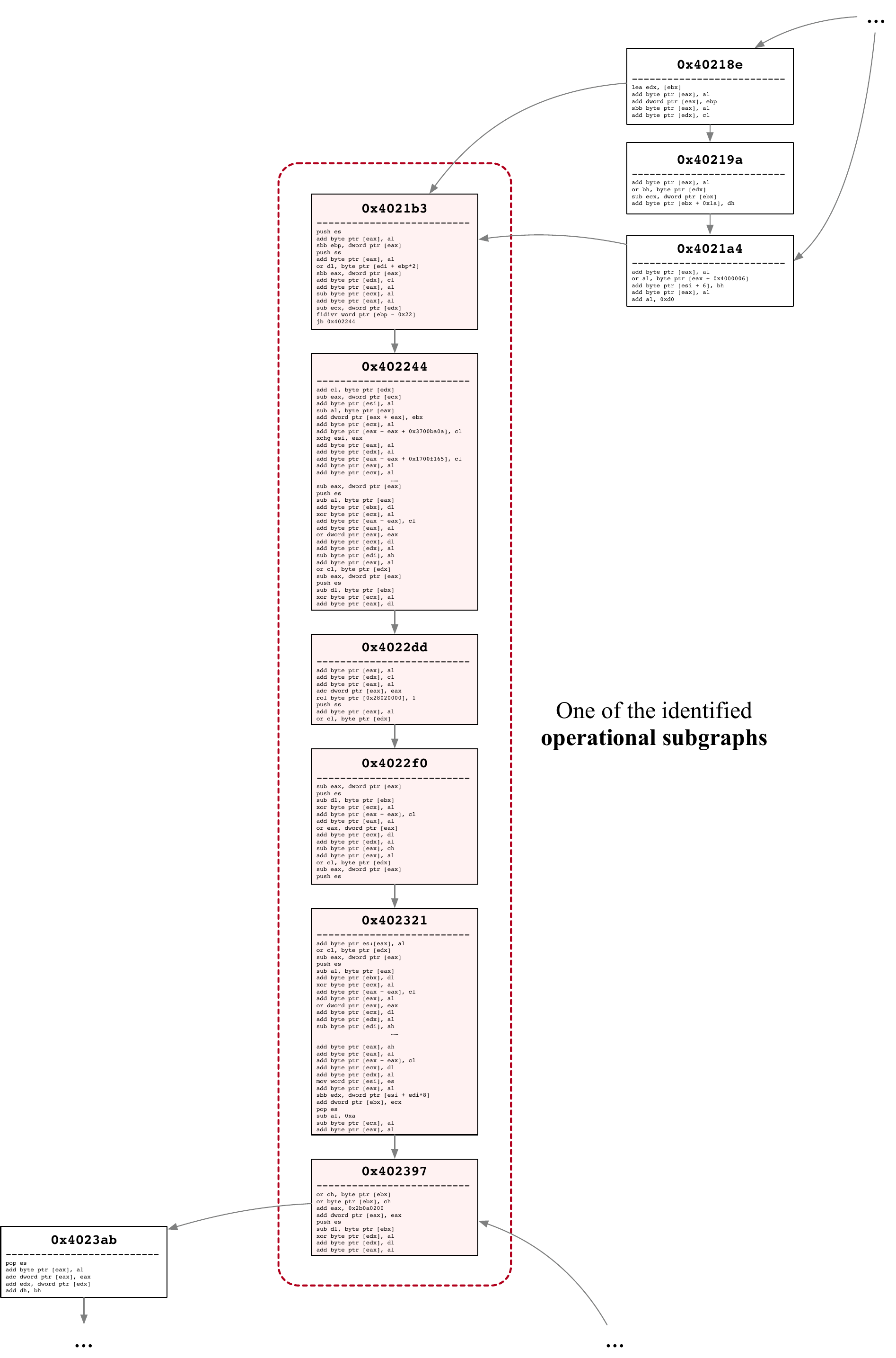}
\vspace{-0.1in}
\caption{
Operational subgraph identified by MalGuard in a case study. The figure shows a partial CFG of the function located at entry address \texttt{0x40210a} in the malware instance \texttt{MemoryDiagnostic.exe}. The red dashed box highlights an operational subgraph identified by MalGuard.
}
\label{fig:case_study}
\end{figure}

The analyzed instance is a prototypical Trojan that deceptively masquerades as a legitimate system diagnostic tool. Its objective is to lurk within an infected system, perform extensive environmental reconnaissance, and subsequently download more destructive payloads such as info-stealers, while potentially propagating to other hosts on the same network.
{We focus on a specific function with the entry address \texttt{0x40210a}, which receives a high importance weight in Equation~\eqref{eq:CGgate} during the inter-function modeling stage of MalGuard, indicating its importance in the overall detection decision.}
Through the {operational role identification} module, MalGuard automatically uncovered a series of operational subgraphs within this function. 
As illustrated in Figure~\ref{fig:case_study}, one identified operational subgraph, highlighted by the red dashed box, forms a long and nearly linear execution chain.

Our manual reverse engineering reveals that this identified operational subgraph corresponds to a customized memory mutation routine, which likely serves as a preparatory decoding stage prior to malicious payload execution.
Importantly, when examined at the level of individual nodes, these basic blocks appear largely benign, as they are dominated by arithmetic operations on obfuscated constants and routine memory updates.
For example, the block at \texttt{0x402244} mainly contains ordinary operations such as subtracting values from memory locations and writing updated bytes back to memory. Similarly, the block at \texttt{0x402321} performs simple arithmetic and byte-level memory updates, including adding a constant to a register and updating memory through instructions. Viewed in isolation, these instructions do not directly indicate a clear malicious action.
However, it is precisely their coordinated execution that gives rise to suspicious behavior.
Across the identified operational subgraph, these blocks repeatedly read values from memory, transform them through arithmetic operations, and write the transformed values back to memory. This coordinated pattern suggests that the program is progressively modifying a memory region, which is consistent with a decoding or unpacking routine used before the actual payload is executed. Such routines are commonly suspicious in malware analysis because they allow malicious code to remain hidden in the file and only reveal its executable form at runtime, making it harder for detectors to identify the harmful behavior before execution. By aggregating these nodes into a coherent operational subgraph, MalGuard is able to capture this latent intent, which would be overlooked by methods focusing solely on local patterns.

This case illustrates that malicious behavior is often not localized within individual nodes but emerges from coordinated execution patterns across multiple nodes in the program graph. By shifting the analytical focus from individual nodes to semantically meaningful operational roles, MalGuard enables more reliable and interpretable malware detection. Such interpretability is particularly valuable in organizational settings, where security analysts require transparent and actionable evidence to support incident response, auditing, and risk assessment.

\subsection{Economic Value Analysis}
\label{sec:5.5}
We further examined the economic value of our proposed method, MalGuard, in the context of organizational cybersecurity.
Modern organizations face increasing exposure to malware attacks, which exploit vulnerabilities in computer systems to steal sensitive data, disrupt critical operations, or paralyze systems for ransom payments.
Such incidents often escalate into large-scale security breaches that incur substantial economic losses.
To mitigate these threats, enterprises often deploy proactive malware detection solutions designed to identify and block malicious software before it can cause harm. Nevertheless, due to inherent limitations in detection capabilities, no method is perfectly effective. Some malware instances are mistakenly classified as benign (i.e., negative) instances and remain active within the system. Accordingly, the risk posed by such undetected malware is quantified by the false negative rate (FNR) of the deployed model~\citep{ye2017survey}. Reducing the FNR is therefore critical to lessening the economic impact of cyberattacks, highlighting both the practical necessity and the economic benefit of more effective detection methods.

More precisely, we let $\text{Cost}_\text{baseline}$ denote the annual malware-related economic loss under the baseline detection method, and $\text{Cost}_\text{ours}$ denote the loss under our proposed MalGuard for the same organizational setting. Assuming that the total cost is approximately proportional to the number of false negatives (i.e., malware mistakenly classified as benign), we have:
\begin{equation} 
\text{Cost}_\text{ours} = \text{Cost}_\text{baseline} \cdot \frac{\text{FNR}_\text{ours}}{\text{FNR}_\text{baseline}}, 
\end{equation}
where $\text{FNR}_\text{baseline}$ is the false negative rate of the baseline model, and $\text{FNR}_\text{ours}$ is that of MalGuard. Consequently, the cost reduction achieved by MalGuard compared to the deployed baseline method is given by:
\begin{align}
\begin{aligned}
\Delta \text{Cost} &= \text{Cost}_\text{baseline} - \text{Cost}_\text{ours}\\
&= \text{Cost}_\text{baseline} \cdot \left(1 - \frac{\text{FNR}_\text{ours}}{\text{FNR}_\text{baseline}} \right). 
\end{aligned}
\end{align}

\begin{table}[t]
    \centering
    \fontsize{10}{10}\selectfont
    \caption{Economic value analysis of malware detection methods. Cost reductions indicate the estimated savings from replacing each baseline method with MalGuard.}
    \label{tab:economic}
    \begin{tabular}{lccc}
        \toprule
        {Method} &
        {FNR} &
        \makecell{Cost Reduction by \\ MalGuard} &
        {Reduction Rate} \\
        \midrule
        MalGuard & 9.77\%  &   -   &   -    \\
        \midrule
        No Detection & 100.00\% & \$2,358,568.89 & 90.23\% \\
        MalConv    & 20.93\% & \$1,393,774.69 & 53.32\% \\
        NonNeg     & 31.32\% & \$1,798,552.54 & 68.81\% \\
        DRSM       & 18.91\% & \$1,263,433.17 & 48.33\%
        \\
        GCN        & 22.33\% & \$1,470,274.84 & 56.25\% \\
        GAT        & 15.35\% & \$950,218.38  & 36.35\% \\
        GIN        & 21.86\% & \$1,445,685.26 & 55.31\% \\
        MAGIC      & 19.53\% & \$1,306,306.79 & 49.97\% \\
        DeepCall   & 15.50\% & \$966,319.03  & 36.97\% \\
        Mal2GCN    & 35.97\% & \$1,903,962.81 & 72.84\% \\
        MalGraph   & 23.41\% & \$1,523,037.39 & 58.27\% \\
        \bottomrule
    \end{tabular}
\end{table}

According to Accenture’s cybercrime study, malware attacks incur an average annual cost of \$2,613,952 per company.\footnote{See \href{https://www.accenture.com/content/dam/accenture/final/a-com-migration/pdf/pdf-96/accenture-2019-cost-of-cybercrime-study-final.pdf}{https://www.accenture.com/content/dam/accenture/final/a-com-migration/pdf/pdf-96/accenture-2019-cost-of-cybercrime-study-final.pdf} (last accessed on Jul 23, 2026).} We thus set $\text{Cost}_\text{baseline}$ to \$2,613,952. Table~\ref{tab:economic} presents the FNRs of different baseline detection methods along with the corresponding relative cost reductions that would be realized if these baseline methods were replaced with MalGuard. In addition, we include a \emph{No Detection} case, representing a situation in which the focal organization adopts no malware detection method (i.e., $\text{FNR}=100\%$). As shown, MalGuard achieves the lowest FNR among all approaches, leading to substantial reductions in economic cost. In particular, compared to the best-performing byte-based baseline (MalConv) and the best-performing  graph-based baseline (DeepCall), our method achieves cost reductions of 53.32\% and 36.97\%, respectively. These results highlight that the superior detection performance of MalGuard contributes to substantial economic value by more effectively detecting malware.

\section{Conclusion}
\subsection{Summary and Contributions}

Malware remains a major cybersecurity threat to organizations.
Although organizations commonly rely on ML malware detection methods to mitigate malware risks, existing approaches still face important limitations.
Byte-based methods learn from raw byte sequences and are therefore vulnerable to evasive behaviors. Graph-based methods provide a more robust alternative by representing software through program graphs, but they often overlook operational roles and fail to learn sufficiently expressive program graph representations for accurate malware detection.
To address these limitations, we propose MalGuard, a novel graph-based malware detection method for enhancing malware risk management. Specifically, we first introduce an operational role identification approach that infers latent roles in CFGs and extracts the corresponding operational subgraphs. Building on the identified operational roles, we further develop a program graph representation learning method that models interactions among operational roles, preserves sparse malicious signals, and captures the hierarchical structure of program graphs for malware detection.
Extensive experiments on a real-world software dataset show that MalGuard outperforms representative ML malware detection methods and reduces the expected cost of undetected malware for organizations.

Our study contributes to the extant literature in two ways. First, our study belongs to the computational design science paradigm in the IS field, which aims to solve important business and societal problems and makes significant methodological contributions to the IS literature~\citep{padmanabhan2022editor, fang2025computational, li2021will}. In this regard, we focus on malware risk management, a critical cybersecurity challenge faced by organizations, and propose a novel graph-based malware detection method to improve organizations’ ability to identify malware before it causes operational and financial damage. Specifically, the key innovations of MalGuard, operational role identification approach and program graph learning method for malware detection, constitute the core methodological contributions of this work. Second, our study contributes to the broad IS cybersecurity literature~\citep{galbreth2010impact, ebrahimi2025radar}. By modeling the characteristics of program graphs and software execution, MalGuard enables organizations to better detect malware, including malware exhibiting evasive behaviors. In doing so, our study extends IS cybersecurity research by demonstrating how MalGuard can be used to support organizational malware risk management and reduce the economic consequences of cyber threats.

\subsection{Implications and Future Work}
Beyond detection performance, our study also offers several practical implications. First, by adopting our proposed method, software users can screen potentially malicious software before it enters their IT infrastructures, thereby substantially reducing the expected costs associated with undetected malware. This capability is particularly important for public-facing and critical-infrastructure organizations, such as financial institutions, transportation operators, and public utility providers. In these cases, a successful malware infection may not only disrupt internal operations but also interrupt essential services, compromise sensitive citizen or customer data, and generate cascading economic and social consequences. Therefore, our method provides organizations with a practical ex ante cost-effective mechanism that helps reduce the broader organizational and societal costs of malware incidents. Second, our study provides implications for cybersecurity governance and policy. Specifically, malware exhibiting evasive behaviors is increasingly common in practice and is intentionally designed to bypass automated detectors while preserving harmful functionality. Cybersecurity governance should therefore move beyond static checklists and encourage the adoption of detection mechanisms that explicitly account for evasive and adaptive threats. Furthermore, our study has broader societal implications. In increasingly digitized societies, public trust in digital technologies is essential for the adoption of IT-enabled services. Malware incidents can undermine this trust by heightening users’ perceived security risks, privacy concerns, and distrust toward digital platforms and institutions. By improving the detection of malicious software, our method helps reduce the likelihood and impact of such incidents, thereby strengthening societal trust in digital environments. This enhanced trust may further promote the diffusion of IT-enabled applications, thereby contributing to social and economic development.

Our study opens up multiple directions for future research. 
First, adapting graph-based malware detection to continuously evolving malware threat presents a valuable future direction. 
In real-world cybersecurity environments, malware authors often develop new evasive behaviors over time, which may reduce the effectiveness of detectors trained on currently observed malware instances. 
Therefore, future studies could extend MalGuard to support continuous adaptation under emerging threats by incorporating concept drift detection~\citep{lu2018learning}, cross-domain learning~\citep{li2021dual}, or adversarial training~\citep{madry2018towards}.
Second, improving the method’s scalability and efficiency for practical enterprise deployment presents another important direction. In real organizational settings, malware detection systems need to scan large volumes of files with minimal delay. More scalable and efficient approaches could further enhance the practical value of the proposed method in organizational cybersecurity operations.

\bibliography{reference}


\clearpage
\renewcommand{\thepage}{A\arabic{page}}
\setcounter{page}{1}
\section*{Appendices}
\begin{APPENDICES}

\section{Complexity Analysis Details}
\label{appendix:complexity}

The time complexity of MalGuard mainly involves three components: identifying operational roles, extracting operational subgraphs, and learning program graph representations for malware detection. Let $|\mathcal{R}|$ denote the number of software instances. For each software instance, let $F$ denote the average number of functions, and $m_{\text{CG}}$ denote the average number of edges in a CG. For each CFG, let $n$ and $m$ denote the average number of nodes and edges, respectively. Let $S$ denote the average number of extracted operational subgraphs per CFG, and let $\widetilde{m}$ denote the average number of edges in the coarsened graph. Let $K$ denote the number of latent operational roles, $d$ denote the hidden dimension, and $L$ denote the number of message passing layers.

\begin{itemize}
    \item \textit{Identifying operational roles.} In Stage 1, MalGuard trains the operational role
    identification module over all CFGs. For each CFG, the inference model requires $\mathcal{O}(md+ndK)$ time to compute the latent role distribution. The reconstruction objective evaluates node-pair relationships in the CFG, requiring $\mathcal{O}(n^2K)$ time. 
    After identifying latent operational roles, MalGuard extracts operational subgraphs by thresholding node-role assignments and finding connected components for each role, which requires $\mathcal{O}(K(n+m))$ time for each CFG. Since operational subgraph extraction is performed only once and is much smaller than the cost of identifying latent operational roles, it can be ignored. 
    Therefore, the per-CFG complexity of Stage 1 is $\mathcal{O}(md+ndK+n^2K)$.
    Across all $|\mathcal{R}|F$ CFGs, the per-epoch training complexity of Stage~1 is $\mathcal{O}\left(|\mathcal{R}|F(md+ndK+n^2K)\right)$. 

    \item \textit{Learning program graph representations.} In Stage 2, MalGuard learns program representations for malware detection. For intra-function modeling, the coarsening layer constructs the coarsened graph structure and computes super-node attributes with a cost of $\mathcal{O}(nd^2)$ for each CFG. The attention-based message passing layers over the coarsened graph require $\mathcal{O}(L(Sd^2+\widetilde{m}d))$ time. The gated pooling layer aggregates super-node representations into a CFG-level representation, with a cost of $\mathcal{O}(Sd)$. This term is lower order relative to the cost of other layers and is therefore omitted from the dominant complexity expression. 
    Since each software instance contains $F$ CFGs, the total intra-function modeling cost for each software instance is $\mathcal{O}(F(nd^2 + L(Sd^2+\widetilde{m}d) ))$.
    For inter-function modeling, message passing over the CG requires $\mathcal{O}(L(Fd^2+m_{\text{CG}}d))$ time, and the gated pooling layer introduces only a lower-order cost of $\mathcal{O}(Fd)$. 
    Therefore, the Stage~2  complexity across $|\mathcal{R}|$ software instances is
$
\mathcal{O}\left(|\mathcal{R}|
\left[
Fnd^2
+
FL(Sd^2+\widetilde{m}d)
+
L(Fd^2+m_{\text{CG}}d)
\right]
\right) = \mathcal{O}\left(|\mathcal{R}|
\left[
Fnd^2
+
FL(Sd^2+\widetilde{m}d)
+
Lm_{\text{CG}}d
\right]
\right),
$
since $S \geq 1$.

\end{itemize}

Summing these components, the overall time complexity of MalGuard per epoch is $\mathcal{O}( |\mathcal{R}| [ F(md+ndK+n^2K) + Fnd^2 + FL(Sd^2+\widetilde{m}d) + Lm_{\text{CG}}d])$.
Since $K$, $d$, and $L$ are fixed hyperparameters, the complexity is mainly determined by the number and size of program graphs. 

\newpage

\section{Detailed Experiment Settings}
\label{appendix:experiment}

We provide the detailed implementation settings and baseline configurations used in our experiments.
All models were implemented in Python 3.8, the implementation details are summarized in Table~\ref{tab:implementation}.
Specifically, MalConv, NonNeg, MAGIC, DeepCall, and Mal2GCN were strictly re-implemented following their papers. For DRSM and MalGraph, we utilized the official implementations released by their authors. The detailed hyperparameter settings of the baselines are summarized in Table~\ref{tab:baseline_config}.

\begin{table}[ht]
    \centering
    \fontsize{8}{8}\selectfont
    \caption{Summary of model implementations.}
    \label{tab:implementation}
    \begin{tabular}{p{4cm}|p{5.5cm}p{5.5cm}}
        \toprule
        \textbf{Operation} & \textbf{Implementation Details}
        & \textbf{Official Link} \\
        
        \midrule
        Coding Environment &
        Python with PyTorch library &
        \href{https://pytorch.org/}{\texttt{https://pytorch.org/}} \\

        \midrule
        MalConv~\citep{raff2018malware}, NonNeg~\citep{fleshman2018non} &
        Re-implemented in PyTorch following their original model architectures  &
        \href{https://pytorch.org/}{\texttt{https://pytorch.org/}} \\

        \midrule
        DRSM~\citep{saha2024drsm} &
        Implemented using the official source code &{\href{https://github.com/ShoumikSaha/DRSM}{\texttt{https://github.com/ShoumikSaha/DRSM}}} \\
        
        \midrule
        GCN~\citep{kipf2017semisupervised}, GAT~\citep{velickovic2018graph}, GIN~\citep{xu2018how} & 
        Implemented with PyG library &{\href{https://pyg.org/}{\texttt{https://pyg.org/}}} \\

        \midrule
        MAGIC~\citep{yan2019classifying}, DeepCall~\citep{chen2023deepcall},
        Mal2GCN~\citep{kargarnovin2024mal2gcn} &
        Re-implemented in PyG following their original model architectures  &
        \href{https://pyg.org/}{\texttt{https://pyg.org/}} \\
        
        \midrule
        MalGraph~\citep{ling2022malgraph}& 
        Implemented using the official source code &{\href{https://github.com/ryderling/MalGraph}{\texttt{https://github.com/ryderling/MalGraph}}} \\
        \bottomrule
    \end{tabular}
\end{table}
\vspace{-0.5cm}

\begin{table}[ht]
    \centering
    \fontsize{8}{8}\selectfont
    \caption{Summary of baseline hyperparameter configurations.}
    \label{tab:baseline_config}
    \begin{tabular}{p{4cm}|p{11.5cm}}
        \toprule
        \textbf{Baseline} & \textbf{Hyperparameter Configurations} \\
        \midrule

        MalConv~\citep{raff2018malware} &
        Maximum input length $D=2$ MB; byte embedding dimension $=8$; number of convolutional filters $=128$; filter width $=500$ bytes; stride $=500$; temporal max pooling. \\

        \midrule
        NonNeg~\citep{fleshman2018non} &
        MalConv backbone; maximum input length $D=2$ MB; final-layer weights constrained to be non-negative; weight decay $=10^{-3}$. \\

        \midrule
        DRSM~\citep{saha2024drsm} &
        MalConv backbone; maximum input length $D=2$ MB; number of ablation windows $n=8$; final prediction obtained by majority voting. \\

        \midrule
        GCN~\citep{kipf2017semisupervised} &
        Two-layer GCN for CFG encoding; two-layer GCN for CG encoding; hidden dimension $=128$; global mean pooling. \\

        \midrule
        GAT~\citep{velickovic2018graph} &
        Two-layer GAT for CFG encoding; two-layer GAT for CG encoding; hidden dimension $=128$; number of attention heads $=4$; global mean pooling. \\

        \midrule
        GIN~\citep{xu2018how} &
        Two-layer GIN for CFG encoding; two-layer GIN for CG encoding; each GIN layer uses a two-layer MLP with ReLU activation; hidden dimension $=128$; global mean pooling. \\

        \midrule
        MAGIC~\citep{yan2019classifying} &
        Two graph convolution layers on the CFG with hidden dimension $=128$; two 1D convolution layers with kernel size $=5$ and padding $=2$; global mean pooling. \\

        \midrule
        DeepCall~\citep{chen2023deepcall} &
        Two graph convolution layers on the CG with hidden dimension $=128$; two 1D convolution layers with kernel size $=5$ and padding $=2$. \\

        \midrule
        Mal2GCN~\citep{kargarnovin2024mal2gcn} &
        Two GCN layers with hidden dimension $=128$; global mean pooling; final-layer weights constrained to be non-negative. \\

        \midrule
        MalGraph~\citep{ling2022malgraph} &
        Two-layer GraphSAGE for CFG encoding; two-layer GraphSAGE for CG encoding; hidden dimension $=128$; dropout rate $=0.2$; global mean pooling. \\

        \bottomrule
    \end{tabular}
\end{table}

\newpage
\section{Extended Experiments on Evasive Techniques}
\label{appendix:evasive}
To further evaluate MalGuard under more challenging conditions, we modify the original malware instances in our test set by injecting evasive behaviors.
Specifically, we employed six different evasion techniques to inject evasive behaviors: FGSM Append~\citep{kreuk2018deceiving}, Slack Append~\citep{suciu2019exploring}, DOS Extension~\citep{demetrio2021adversarial}, DOS Partial Modification~\citep{demetrio2019explaining}, DOS Full Modification~\citep{demetrio2021adversarial}, and Header Field Modification~\citep{nisi2021lost}.
Each technique was applied to the malware instances in the test set, after which models were evaluated on the evasive datasets.
Results show that MalGuard achieves substantial gains over byte-based methods: for example, compared with MalConv, it improves AUC by 25.07\% and AUPRC by 38.99\%, underscoring its strong capability to defend against malware employing evasive behaviors.

\begin{table}[ht]
    \centering
    \fontsize{10}{12}\selectfont
    \caption{Comparison of AUC and AUPRC under different evasion techniques. Values are reported as AUC/AUPRC.}
    \label{tab:evasive}
    \resizebox{\textwidth}{!}{%
    \begin{tabular}{llcccccc}
        \toprule
        \multirow{3.5}{*}{Category} &
        \multirow{3.5}{*}{Method} &
        \multicolumn{6}{c}{Evasive Technique} \\
        \cmidrule(lr){3-8}
        & & \makecell{FGSM \\ Append} 
        & \makecell{Slack \\ Append} 
        & \makecell{DOS \\ Extension} 
        & \makecell{DOS Full \\ Modification} 
        & \makecell{DOS Partial \\ Modification} 
        & \makecell{Header Field \\ Modification} \\
        \midrule
        
        \multirow{3}{*}{Byte-Based}
        & MalConv
        & 67.76/56.51
        & 79.37/67.85
        & 67.85/55.55
        & 72.58/69.93
        & 74.16/68.02
        & 88.76/77.17 \\      
        & NonNeg
        & 72.58/72.31
        & 71.64/71.49
        & 65.18/56.77
        & 78.24/58.06
        & 86.93/79.84
        & 81.64/72.58 \\        
        & DRSM 
        & 77.05/73.45 
        & 80.93/79.92 
        & 79.92/54.88 
        & 80.46/68.24 
        & 82.74/74.49 
        & 90.32/76.63 \\
        \midrule        
        \multirow{4}{*}{Graph-Based}
        & GCN 
        & 89.28/88.21 
        & 90.09/85.86 
        & 89.25/86.77 
        & 89.28/88.21 
        & 84.95/86.02 
        & 90.09/85.86 \\        
        & GAT
        & 91.65/88.04
        & 91.34/88.98
        & 91.11/85.05
        & 90.76/86.51
        & \underline{93.83}/\underline{88.65}
        & 91.34/88.98 \\        
        & GIN
        & 89.42/85.21
        & 91.69/87.20
        & 89.42/85.21
        & 89.46/84.38
        & 89.90/86.13
        & 91.69/87.20 \\        
        & MAGIC
        & 92.06/88.66
        & 91.12/88.02
        & 87.36/84.18
        & 91.19/88.33
        & 91.12/88.02
        & 89.49/85.51 \\        
        & DeepCall
        & 93.72/\underline{92.45}
        & \underline{92.52}/\underline{91.97}
        & \underline{91.94}/\underline{89.27}
        & \underline{92.20}/\underline{89.27}
        & 90.93/88.34
        & \underline{91.94}/\underline{89.27} \\
        & Mal2GCN
        & 82.81/81.36
        & 81.01/80.65
        & 82.48/83.04
        & 83.09/82.23
        & 81.01/80.65
        & 82.21/85.37 \\
        & MalGraph 
        & \underline{93.95}/91.13 
        & 85.14/85.45 
        & 85.28/86.31 
        & 91.29/86.59 
        & 85.73/85.64 
        & 85.28/86.31\\
        \midrule
        
        \multicolumn{2}{c}{MalGuard} 
        & \textbf{95.60}/\textbf{92.54}
        & \textbf{93.83}/\textbf{92.49} 
        & \textbf{93.15}/\textbf{90.76} 
        & \textbf{93.03}/\textbf{91.13} 
        & \textbf{93.89}/\textbf{90.63} 
        & \textbf{93.03}/\textbf{91.13} \\
        \bottomrule
    \end{tabular}
    }
\end{table}

\newpage

\section{Notations}
\label{appendix:notation}
Table~\ref{tab:notation} summarizes the main notations used throughout the paper. Unless otherwise stated, $G=(V,A,X)$ denotes a single CFG under the simplified notation introduced in Section~\ref{sec:3.2}.

\begin{table}[ht]
\centering
\fontsize{10}{12}\selectfont
\caption{Summary of notations.}
\label{tab:notation}
\begin{tabular}{p{0.25\textwidth}p{0.65\textwidth}}
\toprule
\textbf{Notation} & \textbf{Description} \\
\midrule
$\mathcal{R}=\{r_i\}$ & Set of software instances \\
$y_i, \hat{y}_i$ & Ground-truth label and predicted malware probability of $r_i$ \\

$\mathcal{H}=(G_{\text{CG}},\mathcal{G}_{\text{CFG}})$ & Hierarchical program graph of a software instance \\
$G_{\text{CG}}=(V_{\text{CG}},E_{\text{CG}})$ & CG of a software instance \\
$v_m\in V_{\text{CG}}$ & Function node corresponding to the $m$-th function \\
$\mathcal{G}_{\text{CFG}}=\{G_1,\ldots,G_{|V_{\text{CG}}|}\}$ & Set of CFGs associated with function nodes in the CG \\
$G_m=(V_m,A_m,X_m)$ & CFG corresponding to function node $v_m$\\
$G=(V,A,X)$ & Simplified notation for a single CFG when no ambiguity arises \\

$K$ & Number of latent operational roles \\
$\mathbf{z}_i\in\mathbb{R}^K$ & Latent operational role vector of basic block node $v_i$ \\
$\boldsymbol{\mu}_i,\boldsymbol{\sigma}_i$ & Mean and standard deviation vectors of the logit-normal posterior for $z_i$ \\
$\mathcal{OS}=\{\text{OS}_1,\ldots,\text{OS}_S\}$ & Set of operational subgraphs extracted from a CFG \\
$B\in\{0,1\}^{|V|\times S}$ & Assignment matrix between basic block nodes and operational subgraphs \\
$A^{\text{super}},X^{\text{super}}$ & Adjacency matrix and node attribute matrix of the coarsened graph \\

$\mathbf{g}_s^{(l)}$ & Representation of $\text{OS}_s$ at layer $l$ \\
$\alpha_{st}^{(l)}$ & Attention weight from $\text{OS}_t$ to $\text{OS}_s$ at layer $l$ \\
$\boldsymbol{\beta}$ & Fusion gate over operational subgraphs in intra-function modeling \\
$\mathbf{f}$ & CFG-level representation obtained from intra-function modeling \\
$\mathbf{h}_m^{(l)}$ & Representation of function node $v_m$ at layer $l$ \\
$\boldsymbol{\gamma}$ & Fusion gate over function nodes in inter-function modeling \\
$\mathbf{h}_{\text{CG}}$ & Final software representation obtained from inter-function modeling \\
\bottomrule
\end{tabular}
\end{table}

\end{APPENDICES}

\end{document}